\documentclass{aa}  

\usepackage{graphicx}
\usepackage{txfonts}
\usepackage{amssymb}
\usepackage{amsmath}
\usepackage{graphicx}
\usepackage[colorlinks=true, allcolors=blue]{hyperref}
\usepackage{booktabs}
\usepackage{amsfonts}     
\usepackage[T1]{fontenc} 
\usepackage[utf8]{inputenc} 
\usepackage{multicol}
\usepackage{xcolor}
\usepackage[switch]{lineno}
\usepackage{gensymb}
\usepackage{soul}
\usepackage{rotating}
\usepackage{comment}
\usepackage{multirow}
\usepackage{longtable}
\usepackage{lscape}
\usepackage{color}
\usepackage{placeins}

\begin{document}

\title{VLTI-GRAVITY observations of blazars}

\author{Talvikki Hovatta \inst{\ref{FINCA},\ref{MRO},\ref{ELE}} \thanks{\href{mailto:talvikki.hovatta@aalto.fi}{talvikki.hovatta@aalto.fi}} \orcid{0000-0002-2024-8199}, 
Elina Lindfors \inst{\ref{UTU},\ref{FINCA}} \orcid{0000-0002-9155-6199}, 
Heidi Korhonen \inst{\ref{MPIA}} \orcid{0000-0003-0529-1161},
Preeti Kharb \inst{\ref{NCRA}} \orcid{0000-0003-3203-1613},
Markus Wittkowski \inst{\ref{ESOGarching}} \orcid{0000-0002-7952-9550},
Aaron Labdon \inst{\ref{ESOChile}},
Tapio Pursimo   \inst{\ref{NOT},\ref{AU}} \orcid{0000-0002-5578-9219},
Kaj Wiik \inst{\ref{UTU},\ref{FINCA},\ref{MRO},\ref{ELE}}   
}

\institute{Finnish Centre for Astronomy with ESO (FINCA), Quantum, Vesilinnantie 5, University of Turku, FI-20014 Turku, Finland \label{FINCA}
\and
Aalto University Mets\"ahovi Radio Observatory, Mets\"ahovintie 114, FI-02540 Kylm\"al\"a, Finland \label{MRO}
\and
Aalto University Department of Electronics and Nanoengineering, PL~15500, FI-00076 Espoo, Finland\label{ELE}
\and
Department of Physics and Astronomy, University of Turku, FI-20014 Turku, Finland \label{UTU}
\and
Max-Planck-Institut f\"ur Astronomie, K\"onigstuhl 17, D-69117 Heidelberg, Germany \label{MPIA}
\and
National Centre for Radio Astrophysics (NCRA) - Tata Institute of Fundamental Research (TIFR), S. P. Pune University Campus, Ganeshkhind, Pune 411007, Maharashtra, India \label{NCRA},
\and
European Southern Observatory, Karl-Schwarzschild-Str. 2, 85748 Garching bei M\"unchen, Germany \label{ESOGarching}
\and
European Southern Observatory, Alonso de Cordova 3107, Vitacura, Santiago, Chile \label{ESOChile}
\and
Nordic Optical Telescope,  E-38700 Santa Cruz de La Palma, Santa Cruz de Tenerife, Spain  \label{NOT}
\and
Department of Physics and Astronomy, Aarhus University, Munkegade 120, 8000 Aarhus C, Denmark \label{AU}
             }

   \date{Received 15 December 2025; Accepted 17 April 2026}

 
  \abstract
   {}
   {Parsec-scale jets of blazars have so far been spatially resolved only in millimeter and submillimeter wavelengths, where very long baseline interferometry can be used to obtain milliarcsecond-scale images of the jets. We have attempted to spatially resolve the near-infrared emission in jet-dominated blazars for the first time.}
   {We used the VLTI-GRAVITY instrument to obtain milliarcsecond-scale near-infrared interferometric observations of a flaring blazar Ton~599. Additionally, we observed four non-flaring blazars using the GRAVITY-wide mode, where a nearby bright star is used as a fringe tracker.}
   {We modeled the squared visibilities of Ton~599 and found that they are incompatible with a single unresolved point source unless there is a significant amount of additional unknown coherence loss in the instrument. With the present data, we cannot distinguish between a model with an unresolved point source and extended emission or coherence loss and a model with a single Gaussian component. This suggests that we are seeing the unresolved or only partially resolved jet-base in near-infrared wavelengths. The wide-field mode of GRAVITY was challenging for the additional relatively faint targets, resulting in either non-detections or poor-quality data that could not be modeled. }
   {Our observations demonstrate that it is possible to detect the compact jet emission in blazars with near-infrared interferometry, suggesting that with the improved GRAVITY+ instrument it will be possible to spatially resolve and image the near-infrared emission of blazar jets.}


\authorrunning{Hovatta et al.}
   \maketitle
%

\section{Introduction}
Jets are ubiquitous in the Universe and can be seen emerging from
protostars, X-ray binaries, active galactic nuclei (AGNs), and
$\gamma$-ray bursts. Relativistic jets in radio-loud AGNs are thought to
be one of the most efficient particle accelerators, and they shine brightly
at all wavelengths, from the radio to very-high-energy $\gamma$ rays \citep[e.g.,][for reviews]{blandford19,hovatta19}. 
The emission from radio, sometimes up to X-ray energies, is comprised of synchrotron emission of relativistic electrons spiraling in the magnetic field of the jet. Typically, the $\gamma$-ray emission in relativistic jets
is thought to be inverse Compton emission, which requires a compact
emission region close to the black hole, where there are enough seed
photons to be scattered by the relativistic electrons \citep{sikora94}. However, detection of
very-high-energy ($>100$\,GeV) emission in some quasars 
requires the emission to originate further down the parsec-scale jet due
to $\gamma$-$\gamma$ absorption in dense photon fields \citep{tavecchio09}. 

In most high-power jetted AGNs, the jet emission outshines any thermal emission components that could act as the seed photon field for high-energy emission, hindering accurate modeling of the highest energy emission. This is especially true for blazars, where the jet points toward Earth with a small viewing angle, resulting in high Doppler beaming of the jet emission. Spatially resolving the nuclear scales in near-infrared wavelengths would allow for more accurate modeling to be performed. 

Recent advances in near-infrared interferometry, especially with the Very Large Telescope Interferometer (VLTI) GRAVITY instrument \citep{GRAVITY2017}, have allowed the innermost, hot near-infrared-emitting part of the torus in a number of nearby AGNs to be spatially resolved, resulting in a typical size of $< 1$~pc \citep{gravity20, gravity24}.  The main limitation when studying distant jetted AGNs is the lack of suitably bright nearby calibrator stars to enable fringe tracking. However, during major flaring episodes, the near-infrared K-band (2.2 $\mu m$) magnitude of jetted AGNs can reach the limiting magnitude K$<10.5$ for GRAVITY single-field observations.

In this paper, we report the first observations of jet-dominated blazars with VLTI-GRAVITY. We focus on the observations of a flaring blazar, Ton~599, that was observed and detected within our target-of-opportunity program in 2022. Additionally, we report observations of four other blazars that we attempted to observe in the wide-field mode of GRAVITY. Our paper is organized as follows: In Sect. \ref{observations} we describe the GRAVITY observations and auxiliary data, and in Sect. \ref{vismodeling} we describe the analysis of the GRAVITY data for Ton~599. We discuss our results in Sect. \ref{discussion} and end with our conclusions in Sect. \ref{conclusions}.  
   
\section{Observations}\label{observations}
Although blazars are some of the brightest extragalactic sources due to Doppler beaming, their near-infrared emission still rarely reaches the required magnitude for GRAVITY single-field observations. Therefore, between 2018$-$2024 we ran a target-of-opportunity program to trigger GRAVITY observations if a flaring blazar was bright enough for GRAVITY single-field observations, which at that time\footnote{With the improved sensitivity provided by the GRAVITY+ upgrade and especially the installation of the laser guide star mode for all the telescopes, the magnitude limits became significantly higher \citep{gravity_plus_25}.} required a K-band magnitude $<10.5$. In 2022, a new wide-field mode for GRAVITY was introduced, and it became possible to observe a fainter target as long as there was a bright (K$<10.5$) star within $30\arcsec$ of it \citep{gravity_wide_22}. Previously, this distance limit was only $2\arcsec$, making it impossible to find suitable targets for our extragalactic objects. Subsequently, we identified four suitable targets from the MOJAVE (Monitoring Of Jets in Active galactic nuclei with VLBA Experiments) and TANAMI (Tracking AGNs with Austral Milliarcsecond Interferometry) blazar monitoring samples for wide-field observations. In the following, we first describe the observations of a flaring blazar (Sect.~\ref{ton599_observations}) and then our observations of the four targets in the wide-field mode (Sect.~\ref{wide_observations}).

\begin{figure*}[h]
            \includegraphics[width=0.33\textwidth]{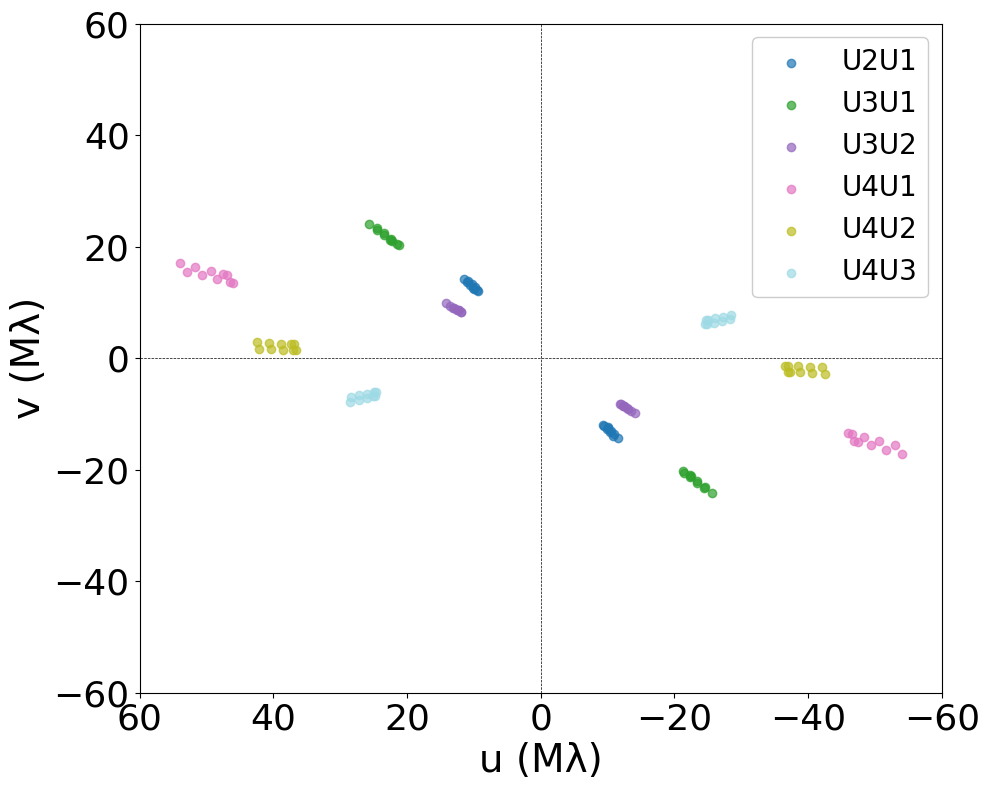}\includegraphics[width=0.33\textwidth]{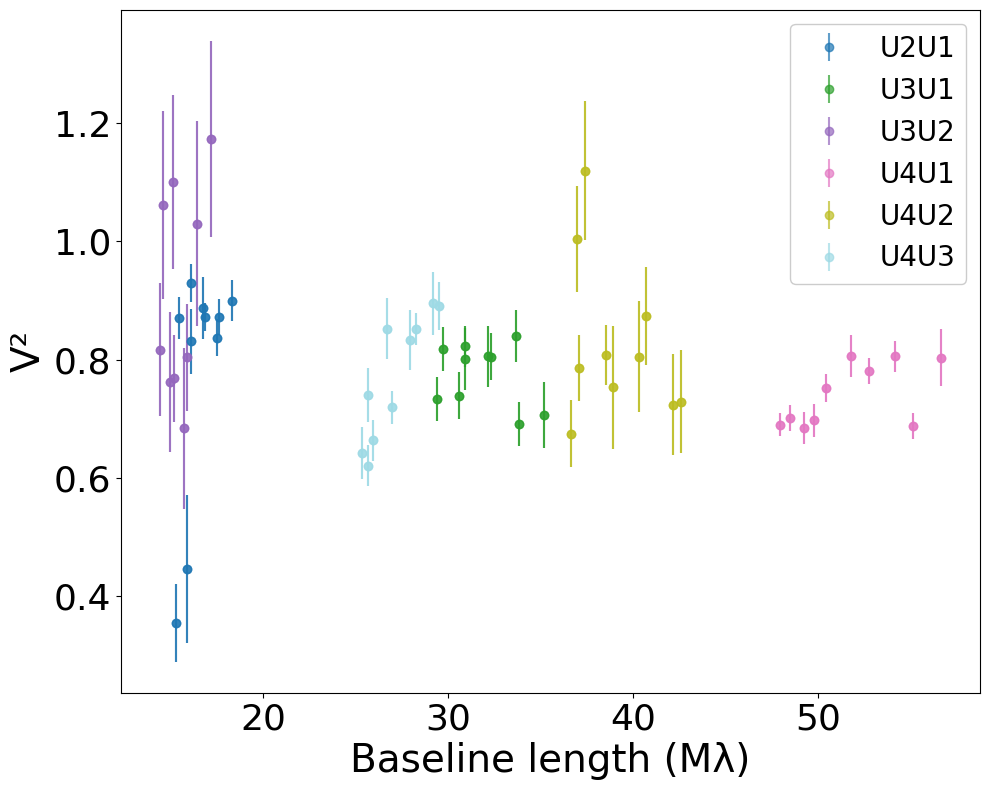}\includegraphics[width=0.33\textwidth]{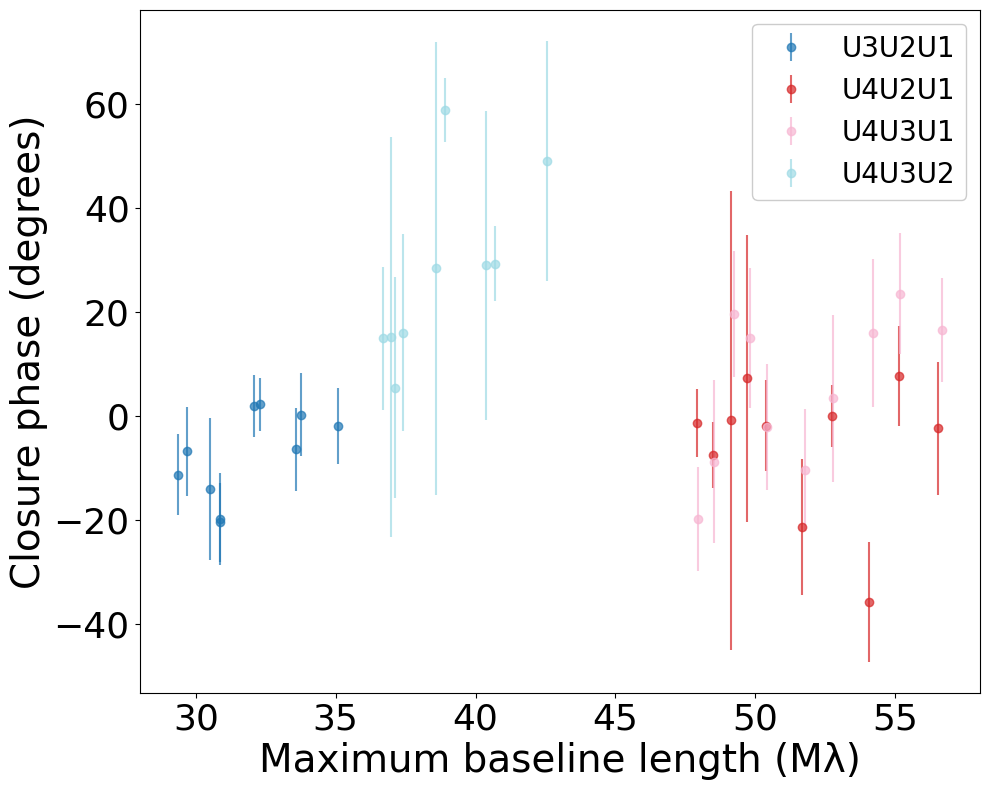}
      \caption{Left: UV coverage of the data of Ton~599. Middle: Squared visibilities as a function of baseline length. Right: Closure phase as a function of maximum baseline of the telescope triangle. In all plots, the colors indicate different telescope pairs or triangles as shown in the legends.
              }
         \label{fig:data}
   \end{figure*}

\subsection{GRAVITY single-field observations of a flaring blazar Ton~599}\label{ton599_observations}
Our sample of blazars was monitored in the optical R-band as part of the Tuorla blazar monitoring program{\footnote{\url{https://tuorlablazar.utu.fi/}} \citep{nilsson18}.
In February 2022, the blazar Ton~599 was seen flaring, reaching an R-band magnitude of $13.88\pm0.02$ on February 16, 2022, 
which led us to request near-infrared observations from the Rapid Eye-Mount telescope (REM) to verify its brightness in the K-band. The source reached the K-band brightness of $10.55\pm0.10$ on February 19, 2022, 
and triggered our target-of-opportunity program.

The VLTI observations using GRAVITY were carried out on February 22, 2022, for a total of 3 hours, starting at 05:22 UT. GRAVITY is an interferometric instrument operating in the near-infrared K band, between 2.0 and 2.4 $\mu$m \citep{GRAVITY2017}. It combines the telescope beams of the four 8 meter VLT telescopes to perform interferometry. The observations were taken in the GRAVITY single-field off-axis mode, where all the light of the target is directed into the fringe tracker and the science fibers are pointed on the empty sky. This enhances the signal-to-noise ratio by a factor of two in observations of faint targets when the light is not split between fringe tracking and science fibers. The resulting spectral resolution is then that of the fringe tracker, where six spectral channels over the band are obtained.

The observations started with an observation of a calibrator star, HD~103846, followed by the science target Ton~599; a calibrator, BD+30~2201; the science target; and again a calibrator, HD~103846. The seeing at the beginning of the observation during the first calibrator star was around $0.9\arcsec$, dropping to $0.55\arcsec-0.65\arcsec$ during the first observation of the science target and increasing again to $0.6\arcsec-0.7\arcsec$ during the second execution of the science target. We first ran a standard pipeline calibration of all the data using the EsoReflex \citep{esoreflex13}  gravity pipeline version 1.7.0 \texttt{gravity\_wkf}, followed by the \texttt{gravity\_viscal} workflow. Upon closer inspection of the calibrator data, the correlated flux during the first calibrator observation of HD~103846 (with worse seeing conditions) was only about half of the correlated flux of the latter observation of HD~103846. This led us to discard the first calibration observation from the final calibration of the data, but we note that this does not affect any of the conclusions presented in the paper.

Because our target brightness was at the limiting magnitude for GRAVITY single-field observations, we followed the procedure outlined in Sect. 2.4 of \cite{gravity20} in our final calibration. As described in that section, the group delay and phase delay values are calculated in real time to monitor the variation of optical path differences between the baselines. The authors noticed that for faint targets, a noticeable fraction of the group delay values exceeded $\gtrsim 3 \mu m$, which resulted in a considerable visibility loss and a reduction of the final visibility squared values. To account for this, we first ran EsoReflex \texttt{gravity\_wkf} and were able to obtain the P2VM-reduced files as output by setting the parameter \texttt{p2vmreduced\_file} to \texttt{TRUE} in the pipeline. We then flagged all frames with the \texttt{GDELAY} parameter of the fringe tracker $\gtrsim 3 \mu m$ by increasing the \texttt{REJECTION\_FLAG} parameter by one bit to turn on the flag.

We then reran the calibration using the EsoRex recipe \texttt{gravity\_vis\_from\_p2vmred} and with the parameter \texttt{use-existing-rejection} set to \texttt{TRUE}. Finally, we used the EsoRex recipe \texttt{gravity\_viscal} to obtain the calibrated closure phase and visibility squared data. Indeed, by applying this additional flagging, the squared visibilities increased significantly from a mean value of 0.56 to a mean value of 0.80. The signal-to-noise ratio of the shortest $2.00\,\mu m$ wavelength bin was very low, and we flagged this bin completely. Additionally, we flagged one point where the uncertainty in the squared visibility was $>0.8$. We also noticed that during the second execution of the target, when the conditions were already getting worse and the source was setting, the squared visibilities were very noisy with several values above one. Therefore, in our final analysis, we only used data from the first execution. These data are shown in Fig.~\ref{fig:data}, where we show the resulting UV coverage of the observations, the squared visibilities as a function of baseline length, and closure phases as a function of maximum baseline length of the given telescope triangle.

\subsection{GRAVITY wide-field observations of four blazars}\label{wide_observations}
To find suitable jet-dominated blazars for observations in the wide-field mode of GRAVITY, we cross-matched the positions of all blazars observed within the MOJAVE \citep{lister18} and TANAMI  \citep{ojha10} very long baseline interferometry (VLBI) monitoring programs with the 
Two Micron All Sky Survey (2MASS) near-infrared catalog \citep{2MASS}. We found four targets that nominally met the observing criteria so that they had a bright star within a $30\arcsec$ distance.

\begin{table*}
\caption{Targets for the GRAVITY wide-field observations.}             \label{table:wide}      
\centering            
\small
\begin{tabular}{l l l l l}        
\hline\hline          
Target name & 2MASS K & Fringe-tracker name & 2MASS K & GRAVITY observing date(s)\\    
\hline     
PMN~J1329$-$5608 & 13.9 & 2MASS J13285936$-$5608191 & 9.9 & Jan 25, 2024\\
PMN~J1604$-$4441 & 13.7 & 2MASS 16042987$-$4441336 & 10.5 & Feb 28, 2024\\
TXS~1811+062 & 14.0 & 2MASS J18133461+0615517   & 8.8 & Jul 3, 2023, Jul 11, 2025, Aug 06, 2025\\
QSO~B2201+1711 & 12.5 & GPM 330.860278+17.426672 & 10.0 & Jul 1, 2023, Jul 6, 2023\\      
\hline                                   
\end{tabular}
\end{table*}

The targets along with their fringe-tracker stars and 2MASS K-band magnitudes are given in Table~\ref{table:wide}, where the dates of the GRAVITY observations are also given. Although the 2MASS magnitudes of all the targets were bright enough, due to the variability of the targets, all of them were fainter during the GRAVITY observations, resulting in very poor-quality data (PMN~J1604$-$4441, TXS~1811+062) or non-detections (PMN~J1329$-$5608, QSO~B2201+1711). 

The case of TXS~1811+062 is interesting because when it was first observed in July 2023, the object was clearly detected, but the observations were interrupted after a short time due to technical problems with one of the telescopes. We therefore requested new observations of the target in the low spectral resolution mode in order to improve the signal-to-noise ratio of the observations. It was observed again in July and August 2025, resulting in a non-detection. This non-detection cannot be explained with different observing conditions or the brightness of the target. As explained in Appendix~\ref{app1811}, this may be due to the shorter coherence length in the low versus medium spectral resolution mode and the uncertainty in the fringe tracker source coordinates.
It is also possible that there were additional instrumental coherence losses in 2025 that are not understood. The GRAVITY-wide mode only provides relative measures instead of absolute visibilities, making it challenging to characterize the amount of coherence losses \citep{gravity_wide_22}.

Details of the observations and the squared visibility amplitudes for all the targets and their fringe-tracker stars are given in Appendix \ref{appendix}. However, due to the poor quality of the data, we did not attempt to model them further.

\section{Modeling the visibility data of Ton~599}\label{vismodeling}
Following \cite{gravity20}, we modeled the squared visibility amplitudes ($V^2$) of Ton~599 in an attempt to resolve the near-infrared emitting region. The effective resolution of an optical interferometer is defined as $\lambda/(2B)$, where $\lambda$ is the wavelength and $B$ is the longest baseline length \citep[e.g.,][]{eisenhauer23}. For GRAVITY, using the unit telescopes, this is approximately 1.6\,mas. However, in \cite{gravity20}, the smallest sizes obtained by fitting the $V^2$ are much smaller, $< 0.5$\,mas, which they consider to be partially resolved.

We used the final calibrated data from Fig.~\ref{fig:data} and the scipy function \texttt{curve\_fit} to perform the fits. We started with the simplest possible model, which is an unresolved component (shown as a solid line in Fig.~\ref{fig:models}). This resulted in a reduced $\chi^2=40.0$ and is clearly not a good fit to the data, as most of the $V^2$ values are below one. 

   \begin{figure}
   \centering
   \includegraphics[width=0.9\hsize]{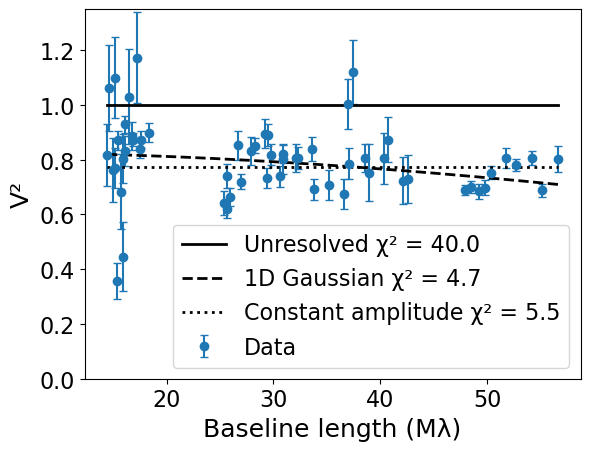}
      \caption{Visibility squared of Ton~599 as a function of baseline length along with the fit models described in the text.
              }
         \label{fig:models}
   \end{figure}

Deviations from one are expected both in the case of extended resolved-out emission and in the case of additional coherence loss due to instrumental effects. 
As discussed in \cite{gravity20}, the $V^2$ values can vary between different executions for the same target, pointing toward additional coherence loss, as the extended emission is not expected to vary on short time scales. This can introduce a scatter of up to 0.2 in the squared visibilities between different observing nights \citep{leftley21}. 

In \cite{gravity20}, the authors therefore modeled the data with a Gaussian component while also allowing the zero-baseline visibility, $V_0^2$, to vary using the function
\begin{equation}
    V^2 = V_0^2exp(-2\pi^2 r_\mathrm{uv}^2\sigma^2),
\end{equation}
where $r_\mathrm{uv}$ is the baseline length in units of wavelength and $\sigma$ is the size of a Gaussian component in radians. The full-width at half maximum (FWHM) of the Gaussian component in milliarcseconds was then obtained from $\mathrm{FWHM}=\sigma2\sqrt{{2ln2}}$ and by converting radians to milliarcseconds. This fit, shown as a dashed line in Fig. ~\ref{fig:models},  gives $\chi^2=4.7$, with the best-fit values of $V_0^2=0.83\pm0.01$ and FWHM$=0.76\pm0.05$\,mas, indicating a partially resolved component that includes 91\% of the observed total flux density.

In \cite{gravity20} and \cite{leftley21}, the authors were able to obtain an estimate of the systematic uncertainty in $V_0^2$ because they had multiple observations of each target taken during different nights. As we only have one observation of a faint target taken under suboptimal conditions, we do not know how much $V_0^2$ is affected by additional coherence losses not accounted for in our calibration. Therefore, we also fit the data with a constant amplitude model: $V^2 = V_0^2$. This fit, shown as a dotted line in Fig. ~\ref{fig:models},  gives $\chi^2=5.5$, with the best-fit value of $V^2=0.77\pm0.01$. This would mean that 88\% of the total flux density is in the compact fully unresolved component.

We used the Akaike information criterion \citep{akaike74} to compare the Gaussian component model with the unresolved constant amplitude component and obtained a difference of $\Delta$AIC$=-1.7$, which does not allow us to differentiate between the models. Our data are thus compatible with an unresolved or at best partially resolved compact component. In addition, there is either additional instrumental coherence loss reducing the $V_0^2$ to $\sim 0.8$ or extended emission on scales larger than about 14\,mas that is resolved out in the interferometric observations.

We note that we also have information on the closure phases (Fig.~\ref{fig:data} right), but they are mostly within $2\sigma$ from zero, indicating that any structure we may see is symmetric within our resolution limits. Our simple fits to $V^2$ only are also consistent with the results obtained from modeling using the \texttt{PMOIRED} python package \citep{merand22}, where the closure phases can easily be included as additional constraints in the models.

\section{Discussion}\label{discussion}
The blazar Ton~599 (also known as 1156+295 or 4C+29.45) frequently flares in wavebands going from radio and optical up to very high-energy gamma-rays \citep[e.g.,][]{ramakrishnan14,hagen-thorn21,adams22}, suggesting a highly dynamic accretion disk and a complex disk-jet connection \citep{Wills1983,Wills1992,Nieppola2007}. Since 2020, the source has been in a very active state with multiple flaring episodes \citep[e.g.,][]{rajput24, manzoor24, sakshi25}, and in February 2022 it reached an optical R-band magnitude of 13.88 (see Sect.~\ref{observations}), triggering our GRAVITY observations. 

During flaring, the near-infrared emission of blazars, including Ton~599, is dominated by synchrotron emission from the jet \citep[e.g.,][]{raiteri14}, and no thermal emission components are visible in the spectral energy distribution (SED; \citealt{MAGIC26}). The disk luminosity in this case can be estimated from the broad line luminosities \citep[e.g.,][]{ghisellini09, punsly11}. The flux of the \ion{Mg}{II} line in Ton~599 is seen to vary with the activity state of the source \citep{Hallum2022}. If we select a value at a high state of the source $F_{\ion{Mg}{II}} = 18.5\times10^{-14}$\,erg\,$s^{-1}$\,$cm^{-2}$, it corresponds to $L_{\ion{Mg}{II}} = 4.6\times10^{44}$\,erg\,$s^{-1}$ at the redshift of 0.725 \citep{Albareti2017} of the source.\footnote{for H$_0$ = 67.8 km~s$^{-1}$~Mpc$^{-1}$, $\Omega_{m}$ = 0.308, $\Omega_{v}$ = 0.692} Using the scaling between the bolometric luminosity and the \ion{Mg}{II} line luminosity from \cite{punsly11}, we obtained a disk luminosity of $L_d = 1.4\times10^{47}$\,erg\,$s^{-1}$.

For simplicity, it is often assumed that the torus reprocesses all the emission from the accretion disk, and the luminosity of the torus then corresponds to a single-temperature blackbody-emitting dust with a luminosity equal to the disk luminosity \citep[e.g.,][]{ghisellini09}. Typically, it is assumed that near-infrared emission from the torus is produced by hot dust at a temperature of about $1200-1500$\,K, which is in line with early observations of quasars \citep{barvainis87}. 
We can therefore compare the brightness of the compact emission we observed in Ton~599 to the expected brightness of such a torus emission. We can estimate the brightness of the compact component using our K-band REM observations taken a night before on February 21, 2022, where its K-band magnitude was $10.69\pm0.09$, resulting in a flux density of $35\pm4$\,mJy. The partially resolved or unresolved component includes $\sim90$\% of the total flux density, which is $\sim31.5$\,mJy. Figure~\ref{fig:torus} shows the observed value compared to a single-temperature blackbody emission from hot dust at a temperature of 1200\,K,  corresponding to silicate dust grains, and at a temperature of 1500\,K, corresponding to graphite dust grains \citep[e.g.,][]{Honig2019}.} The required disk luminosity would have to be five to ten times larger to explain the compact emission we see, making the torus origin unlikely.

\begin{figure}
    \centering
    \includegraphics[width=1.0\linewidth]{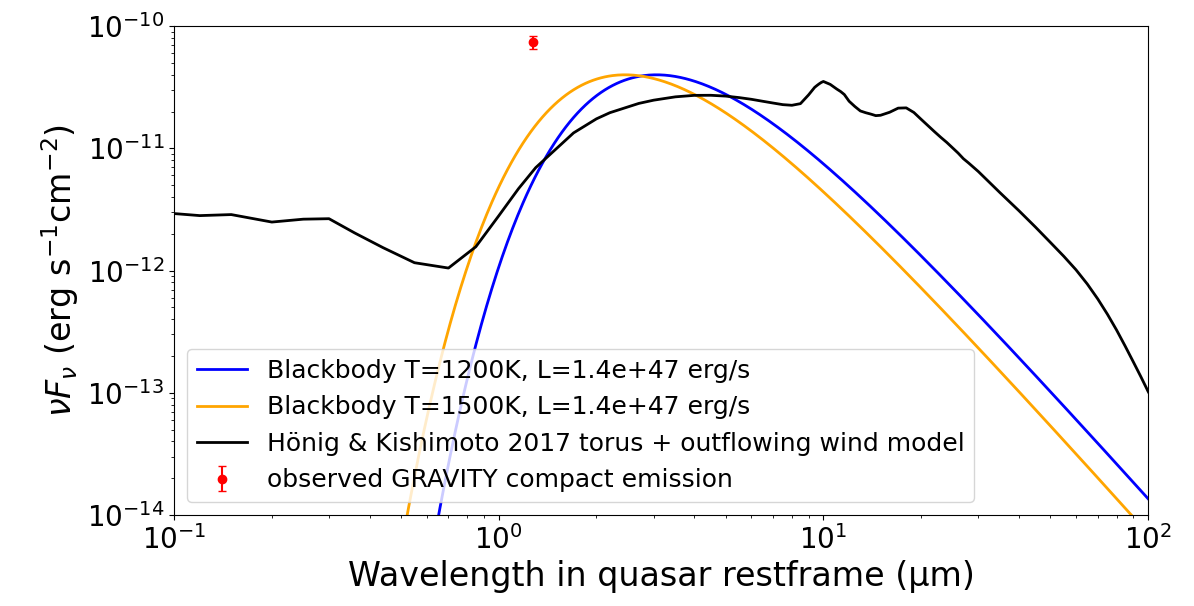}
    \caption{Spectral energy distribution of a single blackbody component with a  luminosity corresponding to the disk luminosity of Ton~599 and temperatures of 1200\,K (blue) and 1500\,K (orange) along with a model for a torus including an outflowing wind from \citet[black]{honig17}. The observed value for the compact emission in our GRAVITY data is shown by the red symbol.}
    \label{fig:torus}
\end{figure}

\begin{figure*}[h]
\centering
\includegraphics[width=0.8\textwidth]{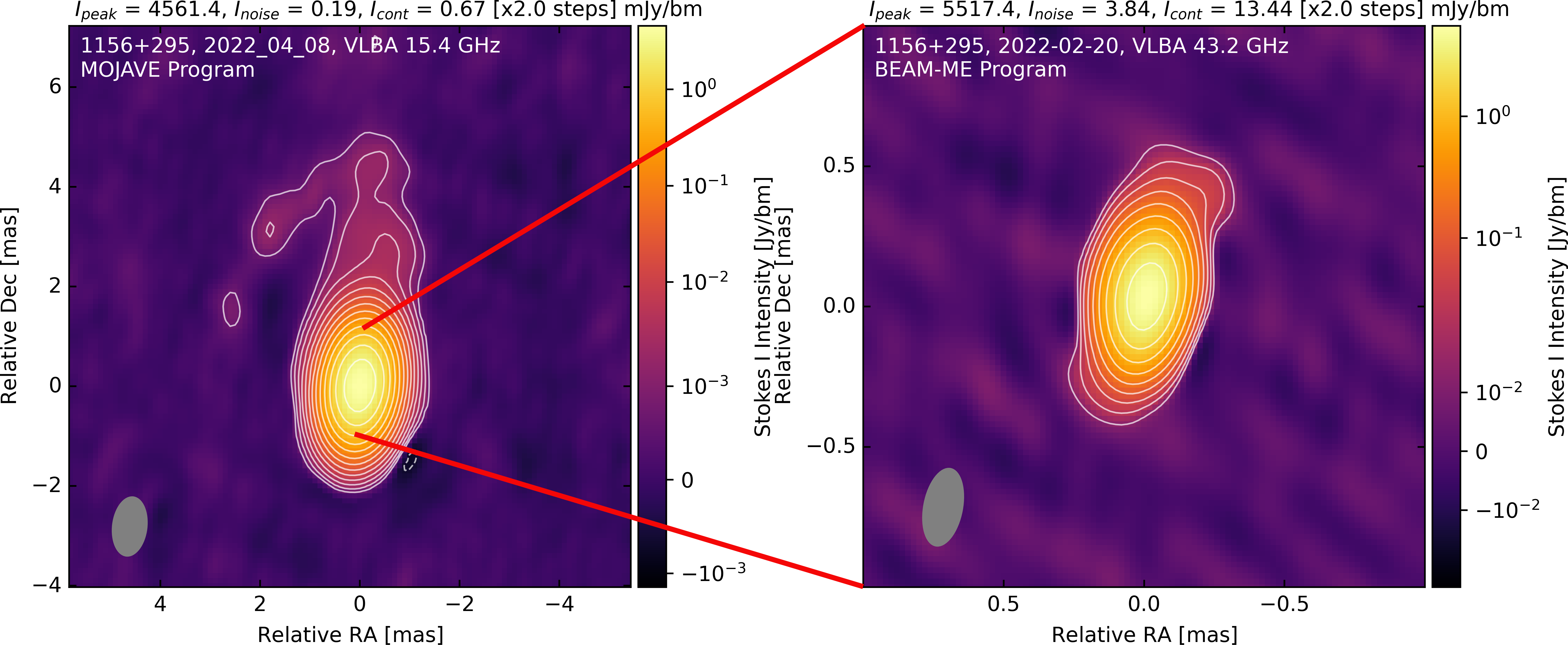}
\caption{Left: Image at 15.4\,GHz of Ton~599 observed by the VLBA within the MOJAVE Program in April 2022. The relativistic jet extends toward the north. Right: Image at 43.2\,GHz from VLBA as part of the BEAM-ME Program on February 20, 2022, two days before our GRAVITY observations. A compact jet with a size of $< 1$~milliarcsecond can be seen.}\label{fig:VLBA}
\end{figure*}

We note that this conclusion would not change even if more complex clumpy or outflowing torus models  with a more realistic dust composition \citep[e.g.,][]{nenkova08,honig17} were used, as these mainly affect the mid-infrared luminosity of the torus. We show one example from the suite of models in \cite{honig17} in Fig.~\ref{fig:torus}, where we have selected the models for an object at an inclination of $0\deg$ to match the inclination of Ton~599 (see below) and scaled it according to the sublimation radius given for the model and the disk luminosity we obtained while also accounting for the redshift of the source. We then selected the model that gives the highest flux at 2.2\,$\mu$m, but as expected, the difference to the simple blackbody model is most pronounced in the mid-infrared range and cannot explain our observations.

Moreover, it is unlikely that the inner torus is as luminous as we have estimated based on the \ion{Mg}{II} flux because during lower activity states of the source, it should become clearly visible in the SED, which is not the case \citep{raiteri14}. For example, in 2017 when the source was first detected at very-high gamma-ray energies by the MAGIC telescopes, its flux was ten times lower than we observed, but no signs of thermal dust emission are seen in the SED \citep{MAGIC26}. The authors used a lower flux for the \ion{Mg}{II} line and a different scaling relation between the line luminosity and disk luminosity and obtained a disk luminosity that is about 40 times lower than our estimate from the high flux state. \cite{Malmrose2011} detected the hot dust emission in the SED of two blazars, which was best fit with a blackbody at a temperature of $\sim1200$\,K. They found the dust luminosity to be $\sim7-8\times10^{45}$\,erg\,$s^{-1}$, which is roughly 20 times lower than our value. Therefore, our disk luminosity is likely an upper limit, supporting the nonthermal origin of the compact emission we observed with GRAVITY.

Such a high disk luminosity would also result in a large inner hot torus radius. Using the scaling relation from \cite{ghisellini09}, which essentially corresponds to the scaling from \cite{barvainis87} for dust at a temperature of 1200\,K, we would obtain a radius of 9\,pc. This is much larger than what is seen in the nearby jetted quasar 3C~273 and the radio galaxy 3C~120, where the near-infrared emission was found to be consistent with emission from a hot dust ring with a size $\lesssim 0.5$\,pc \citep{gravity20}. If we use the scaling relation between the bolometric luminosity of the torus and its radius obtained using GRAVITY data of nearby AGNs from \cite{gravity24}, we obtain a smaller size of 2.3\,pc, which would still be much larger than what is observed in the nearby AGNs. 

Ton~599 is a member of the original MOJAVE sample \citep{Lister2009}. A maximum apparent jet speed of $24.6\pm2.0$\,c was observed through multi-epoch MOJAVE observations \citep{lister19}. From the apparent motions and the timescale of the observed decline in flux of VLBI knots from 2008 to 2013, \citet{Jorstad2017} derived a Doppler factor of $\delta=12\pm3$, a Lorentz factor of $\Gamma=10\pm3$, and a jet viewing angle of $\Theta\le2.5\degr$. 

The VLBI images of Ton~599 reveal a core-jet structure with an oscillating jet on milliarcsecond scales that aligns with the arcsecond-scale jet at a distance of several tens of milliarcseconds from the core \citep{Hong2004}. The arcsecond-scale jet and lobe are also S-shaped, with the northern jet hosting a ``hotspot'' \citep{Antonucci1985, Kharb2010}. The 
northern jet corresponds to the relativistically beamed VLBI jet that is oriented at an angle close to our line of sight \citep[$\le2.5\degr$;][]{Hallum2022}. Ton~599 was also observed in milliarcsecond resolution at 43\,GHz (7\,mm wavelength) with the Very Long Baseline Array (VLBA) on February 20, 2022, just two days before our observations. The 43.2\,GHz VLBA image shown in Fig.~\ref{fig:VLBA} was obtained within the BEAM-ME monitoring program \citep{Jorstad2017, weaver22}. The data were calibrated and imaged as described in \cite{weaver22}, and the fully calibrated data along with the CLEAN model are provided on the program website.\footnote{\url{https://www.bu.edu/blazars/BEAM-ME.html}} We input the data and the model into Difmap \citep{shepherd97} and wrote out the image in the fits format for further plotting in python. 

The 43\,GHz image shows a compact jet with an extent of $< 1$~mas, which is consistent with the compact unresolved component seen in the GRAVITY data, indicating that we are seeing synchrotron emission from the unresolved jet base. In Fig. 2 (left), we show a higher-sensitivity 15.4\,GHz radio image of the source taken in April 2022 within the MOJAVE monitoring program \citep{lister18}. The data were calibrated and imaged as described in \cite{lister18}, and the fully calibrated images are provided on the program website.\footnote{\url{https://www.cv.nrao.edu/MOJAVE/}} 
The 15.4\,GHz image shows how the relativistic jet in Ton~599 extends toward the north. However, the UV coverage of our GRAVITY data, shown on the left-most panel in Fig.~\ref{fig:data}, illustrates that in these observations, we only sampled the east-west structures in the source, and therefore it would be challenging to resolve the jet with the current GRAVITY data, which may explain why we only see an unresolved component.

If the zero-baseline visibility level in the data is correct and not affected by instrumental effects, then $\sim10$\% of the flux in the source must originate from scales larger than about 14~mas, which is the largest scale our observations probe. This size corresponds to about 100~pc at the source redshift. The flux density of the extended component ($\sim4$\,mJy) is much brighter than the luminosity of the host galaxy of the source \citep{olguin-iglesias16}, and some other origin for it should be sought. 

It is possible that the emission is a larger-scale emission from the jet itself. Additionally, some extended emission could possibly be seen from winds or sheaths around the jet. For example, it has been predicted that strong winds are launched from hot accretion flows \citep[e.g.,][]{blandford99, narayan12, yuan12}, which according to recent simulations can reach up to 15\% of the energy flux of the jet \citep{yang21}. Indeed, such winds have been detected in nearby low-luminosity AGNs \citep[e.g.,][]{Gallimore2024,shi25} as well as in powerful radio galaxies in HI absorption studies \citep{Schulz2021}. However, Ton~599 is a flat-spectrum radio quasar, and such objects are assumed to host thin disks. Therefore it is unclear whether such a strong wind would be present.

\section{Conclusions}\label{conclusions}
We observed the flat-spectrum radio quasar Ton~599 for the first time in milliarcsecond-scale resolution in near-infrared wavelengths with the GRAVITY instrument. The observations were conducted when the source was flaring and reached a K-band magnitude of 10.5. The source was clearly detected, demonstrating the feasibility to observe blazars with GRAVITY. We modeled the squared visibilities as a function of the baseline length but could not distinguish between a compact unresolved component and a partially resolved Gaussian component. In both cases there must be either additional instrumental coherence loss or extended emission on scales larger than 14\,mas that is resolved out in the observations to explain the visibility amplitudes. With the present data, we cannot distinguish between the two alternatives. We can rule out the inner region of a hot dusty torus as the  sole origin of the compact emission and suggest that we are observing synchrotron emission from the unresolved jet base. 

Additionally, we reported on the observations of four other non-flaring blazars taken in the GRAVITY wide-field mode. The challenging observations of these faint targets resulted in either non-detections or poor-quality data that could not be further modeled.

\section*{Data availability}
The calibrated squared visibilities of Ton~599 are available as a table vis2.dat in electronic form at the CDS via anonymous ftp to \url{cdsarc.u-strasbg.fr (130.79.128.5)} or via \url{http://cdsweb.u-strasbg.fr/cgi-bin/qcat?J/A+A/}. 

\begin{acknowledgements}
We thank the referee for the useful comments that helped to improve the clarity of the paper. TH acknowledges support from the Research Council of Finland projects 317383, 320085, 345899, and 362571 and from the European Union ERC-2024-COG - PARTICLES - 101169986. Views and opinions expressed are however those of the author(s) only and do not necessarily reflect those of the European Union or the European Research Council Executive Agency. Neither the European Union nor the granting authority can be held responsible for them. EL was supported by the Research Council of Finland projects 317636, 320045, and 346071. HK acknowledges the financial support from European Southern Observatory's Science Support Discretionary Fund (SSDF) and the Visitor and Mobility program of the Finnish Centre for Astronomy with ESO (FINCA). PK acknowledges the support of the Department of Atomic Energy, Government of India, under the project 12-R\&D-TFR-5.02-0700 and the Visitor and Mobility program of the Finnish Centre for Astronomy with ESO (FINCA).
Based on observations collected at the European Southern Observatory under ESO programmes 108.221C.001, 111.24L9.001, 112.25FL.001, and 113.26E7.001. We thank the director of the REM telescope Emilio Molinari for approving our DDT request (proposal 44908) and Dino Fugazza for the swift scheduling of our observations. Tuorla blazar monitoring program makes use of data from the Joan Oró Telescope (TJO) of the Montsec Observatory (OdM), which is owned by the Catalan Government and operated by the Institute for Space Studies of Catalonia (IEEC).
This study makes use of VLBA data from the VLBA-BU Blazar Monitoring Program (BEAM-ME and VLBA-BU-BLAZAR; http://www.bu.edu/blazars/BEAM-ME.html), funded by NASA through the Fermi Guest Investigator Program. The VLBA is an instrument of the National Radio Astronomy Observatory. The National Radio Astronomy Observatory is a facility of the National Science Foundation operated by Associated Universities, Inc. This research has made use of data from the MOJAVE database that is maintained by the MOJAVE team (Lister et al. 2018). This publication makes use of data products from the Two Micron All Sky Survey, which is a joint project of the University of Massachusetts and the Infrared Processing and Analysis Center/California Institute of Technology, funded by the National Aeronautics and Space Administration and the National Science Foundation.
\end{acknowledgements}

\bibliographystyle{aa} 
\bibliography{gravity_blazars} 

\begin{appendix}
\onecolumn

\section{GRAVITY-wide observations of four blazars}\label{appendix}
The data of our wide-field observations described in Table~\ref{table:wide} were calibrated using EsoReflex \texttt{gravity\_wkf} workflow and the gravity pipeline version 1.7.0. The data were mostly taken in the medium spectral resolution mode, apart from the two executions of TXS~1811+062 in 2025 that were taken in the low spectral resolution mode.

Here we show the UV coverage, visibility squared, and closure phase plots for all the targets and their fringe tracker stars. We did not apply any additional flagging on these data as none of the targets had good-quality data for modeling. Therefore the visibility squared values in the shortest and longest wavelength bins, which correspond to the longest and shortest baseline lengths in any given telescope pair, show typically lower values due to the throughput of the instrument being worse at the band edges \citep{gravity24}. This is most clearly seen in the data of the fringe tracker sources.

\subsection{PMN~J1329$-$5608}

\begin{figure*}[h]
            \includegraphics[width=0.33\textwidth]{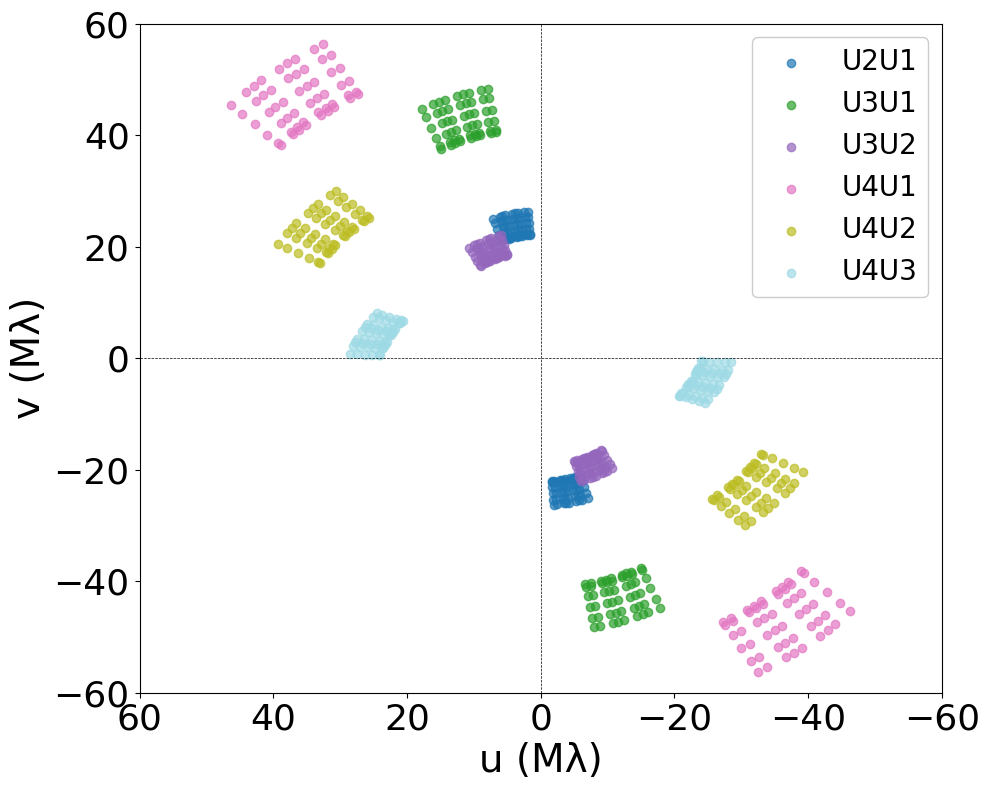}\includegraphics[width=0.33\textwidth]{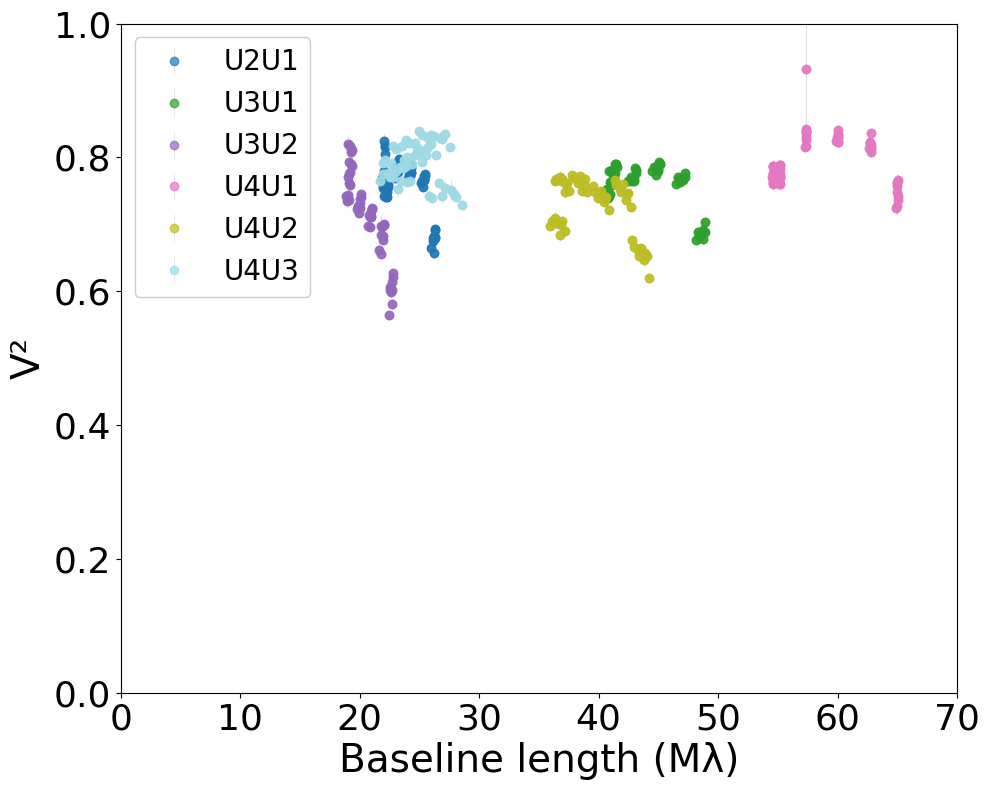}\includegraphics[width=0.33\textwidth]{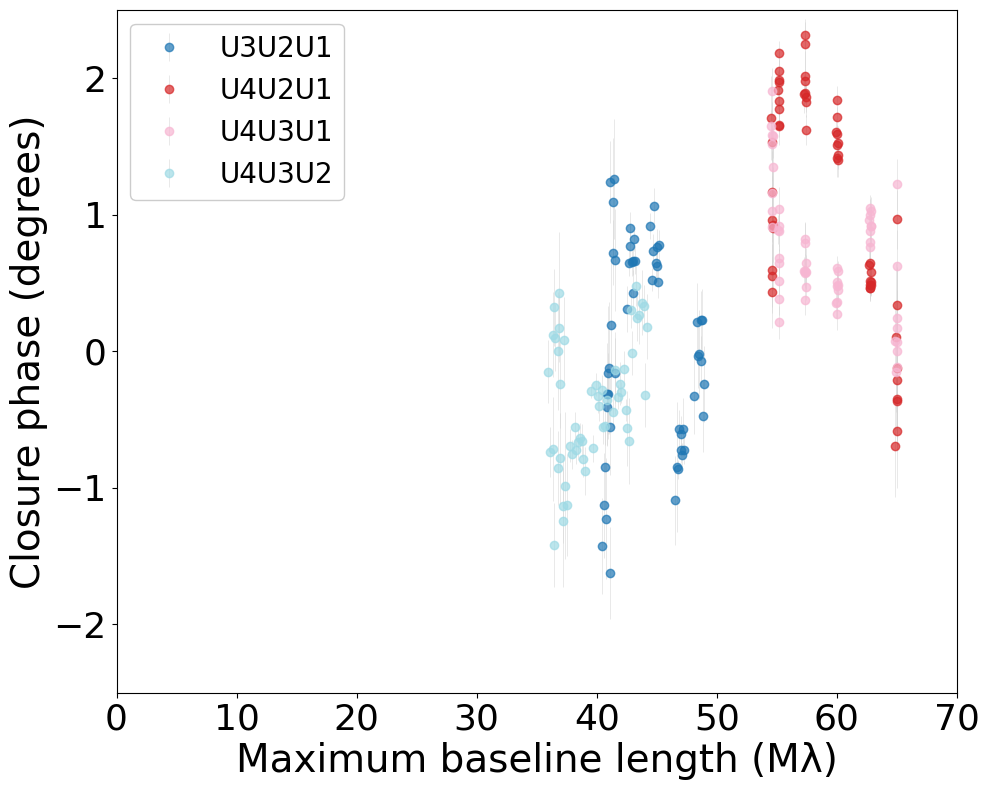}\\
            \includegraphics[width=0.33\textwidth]{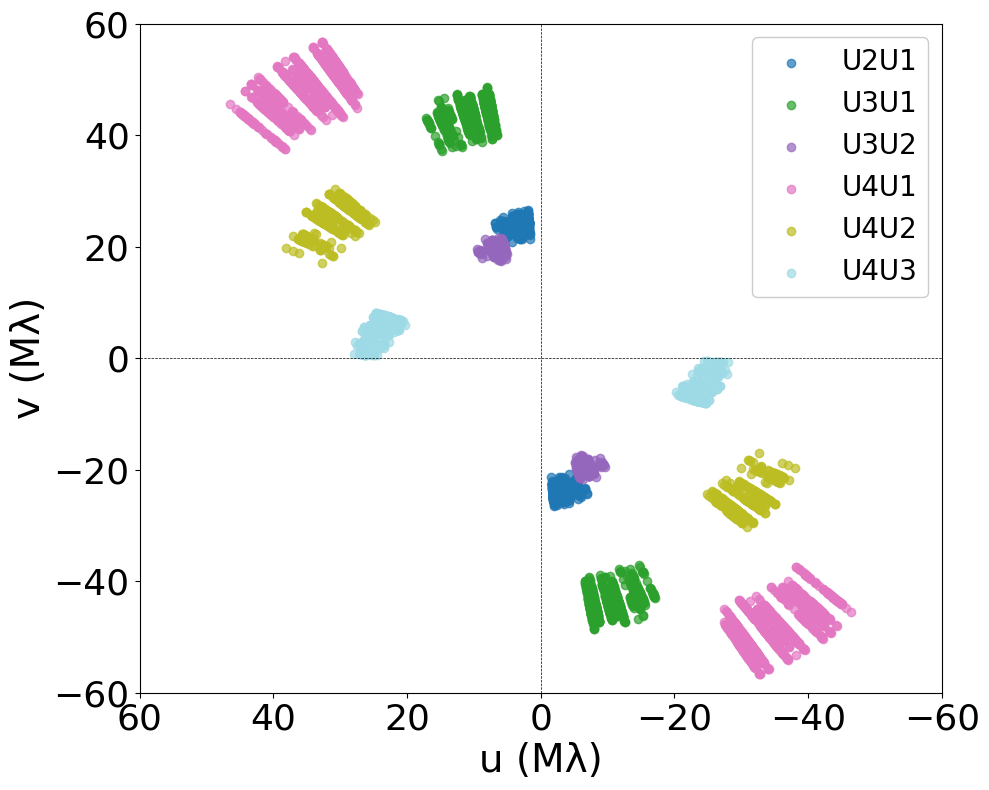}\includegraphics[width=0.33\textwidth]{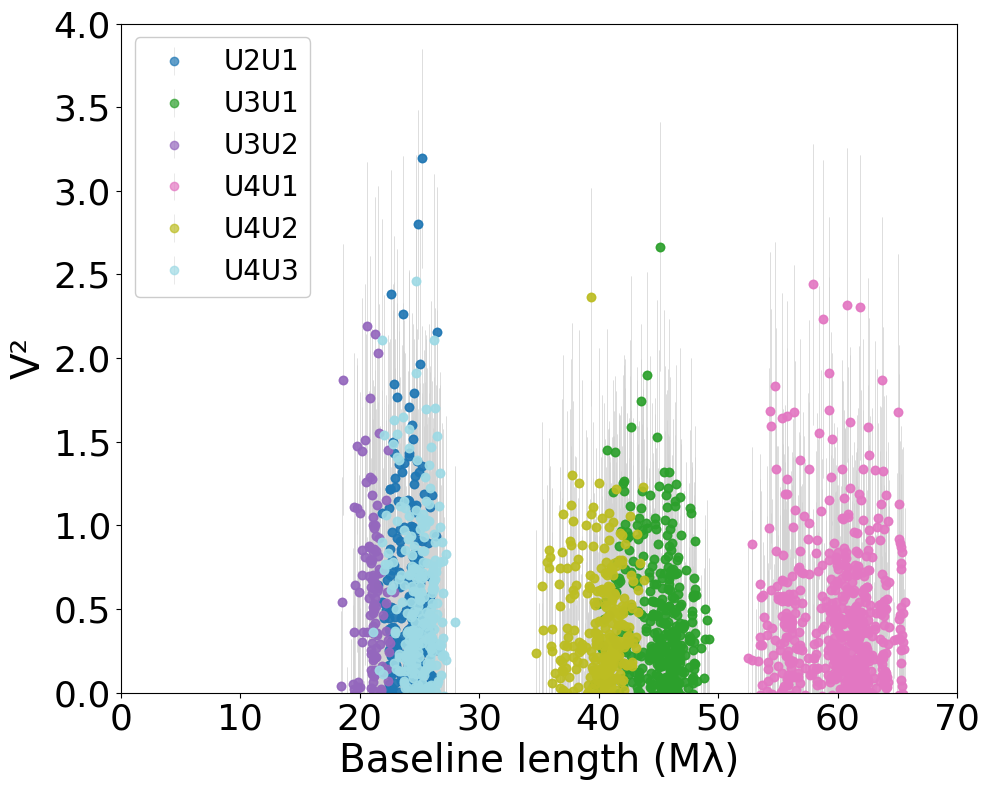}\includegraphics[width=0.33\textwidth]{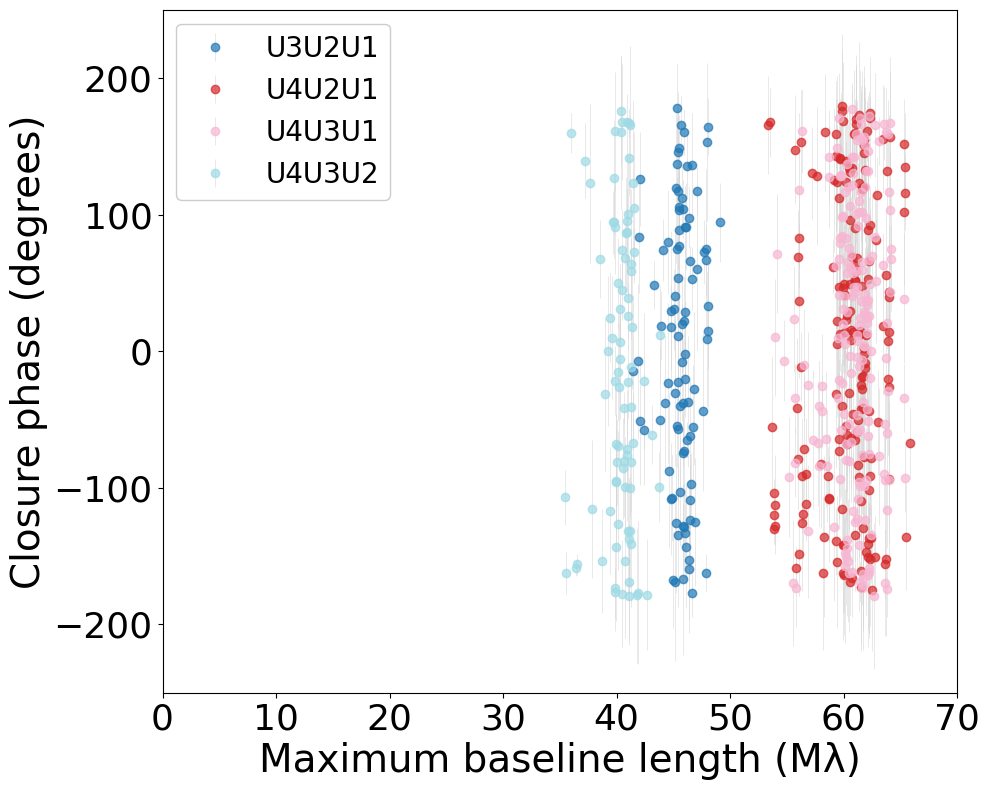}
      \caption{Top: Data of the fringe tracker 2MASS~J13285936$-$5608191. Bottom: Data of the target PMN~J1329$-$5608.
      Left: UV coverage. Middle: Squared visibilities as a function of baseline length. Right: Closure phase as a function of maximum baseline of the telescope triangle. In all plots, the colors indicate different telescope pairs or triangles as shown in the legends.
              }
         \label{fig:data_J1329}
   \end{figure*}

PMN~J1329$-$5608 is a BL~Lac type object observed in the radio band within the TANAMI program \citep{muller18}. We observed it in January 2024 along with a fringe tracker source 2MASS~J13285936$-$5608191 selected from the 2MASS catalog. The separation between the fringe tracker and the target was 22$\arcsec$. Although the fringe tracker is clearly detected (Fig.~\ref{fig:data_J1329}), the target itself shows very low squared visibility amplitudes with large uncertainties and closure phases corresponding to noise and we deem it as a non-detection.

\subsection{PMN J1604$-$4441}
\begin{figure*}[h]
            \includegraphics[width=0.33\textwidth]{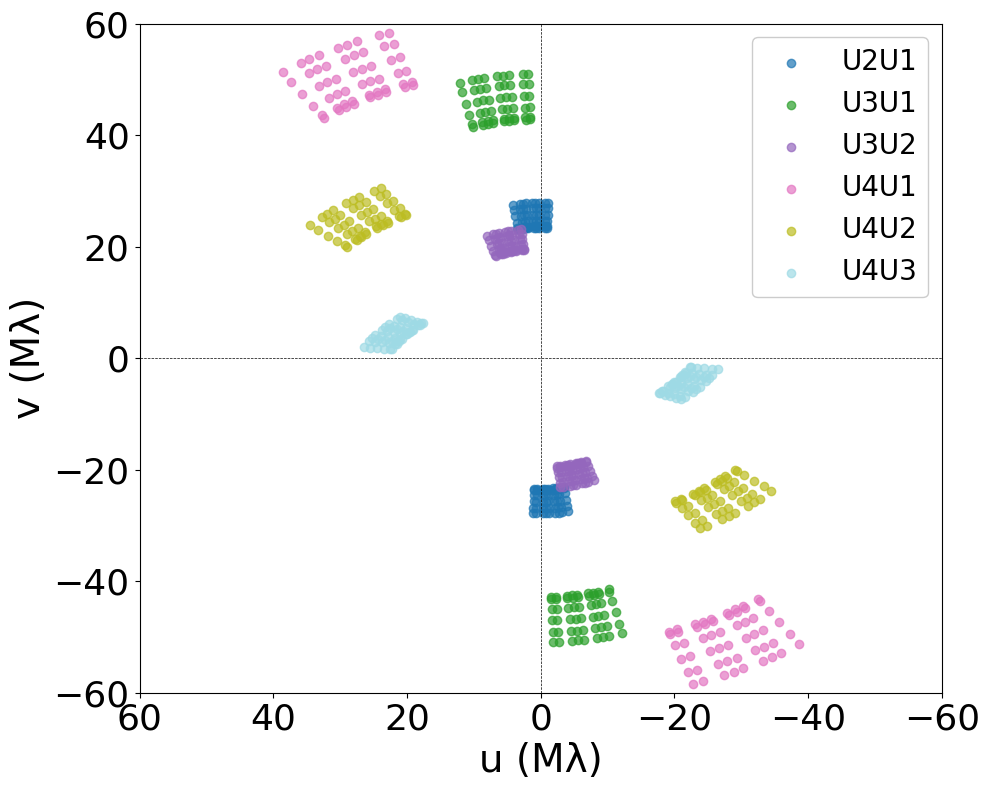}\includegraphics[width=0.33\textwidth]{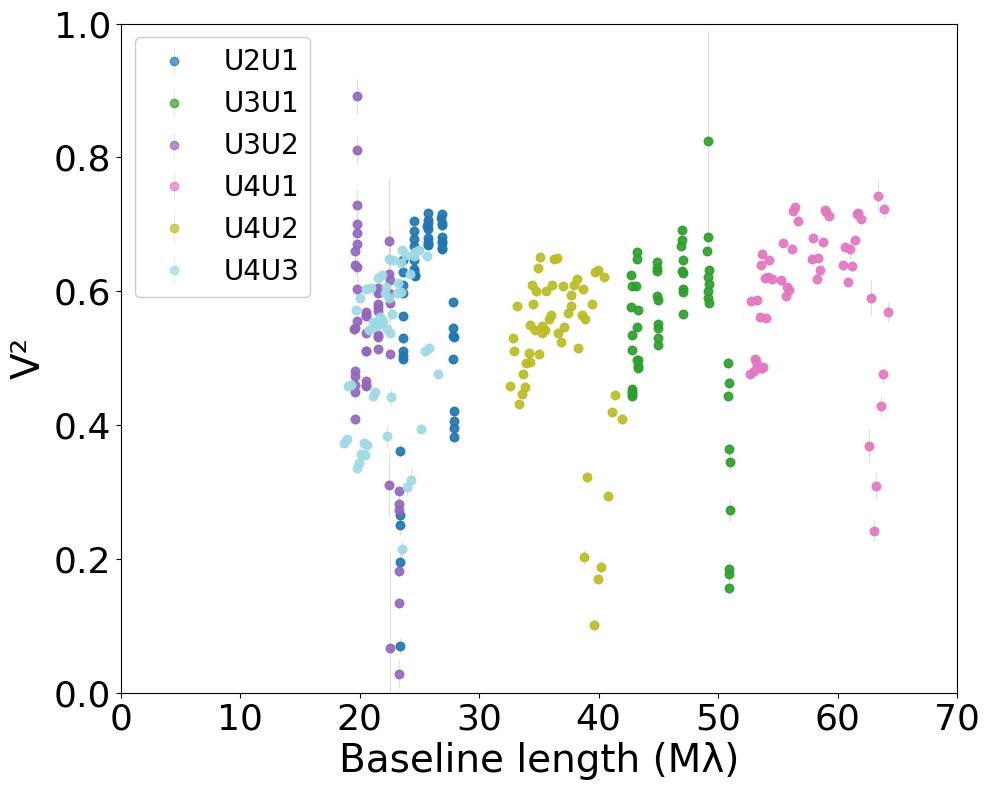}\includegraphics[width=0.33\textwidth]{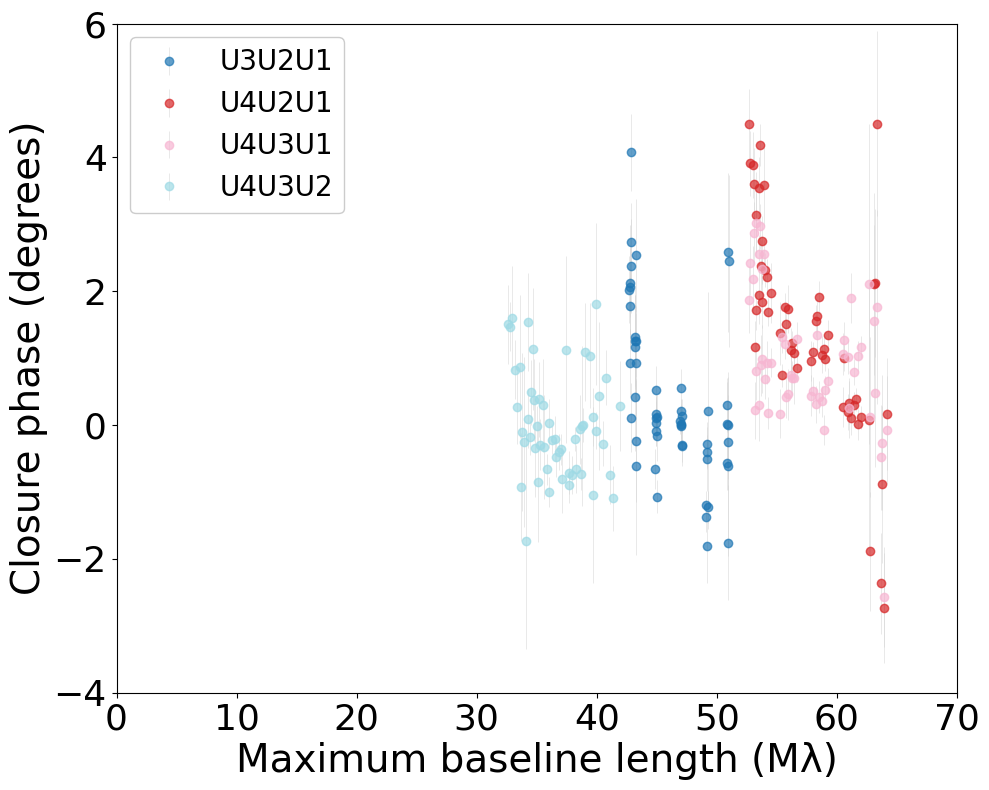}\\
            \includegraphics[width=0.33\textwidth]{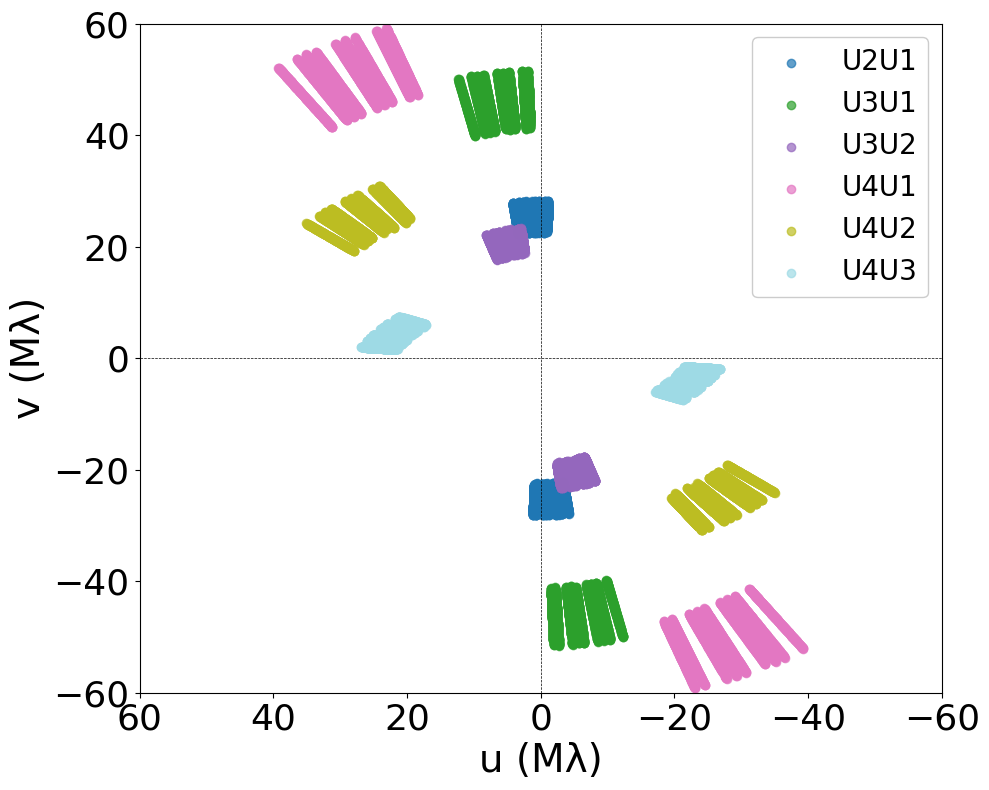}\includegraphics[width=0.33\textwidth]{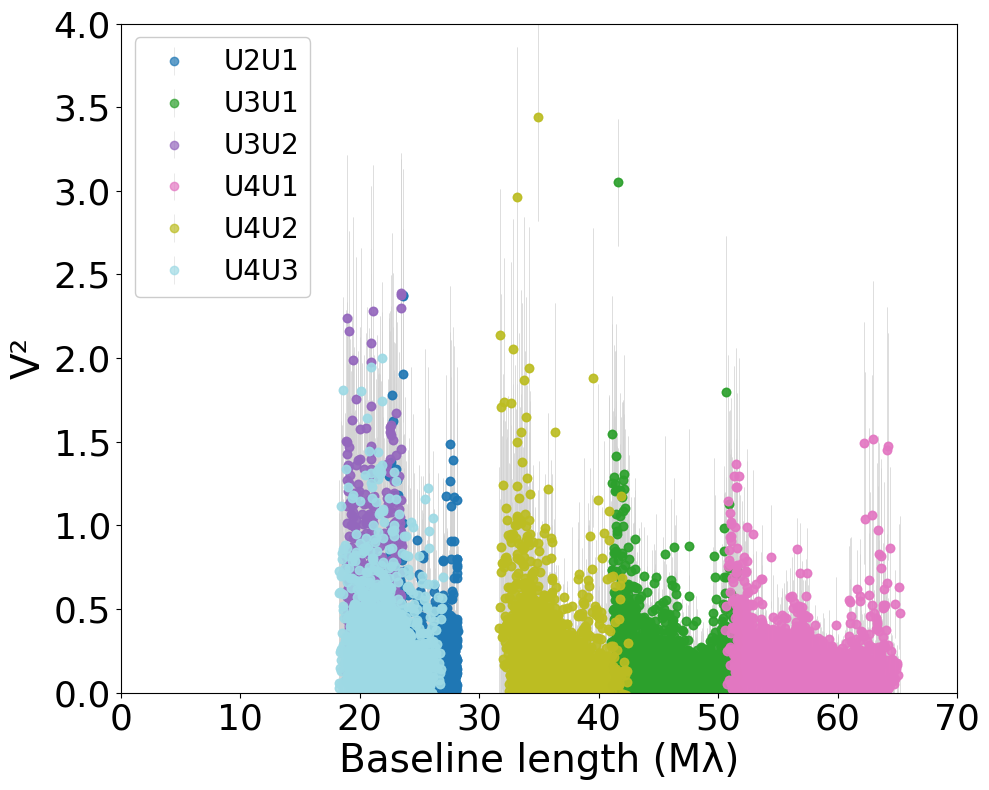}\includegraphics[width=0.33\textwidth]{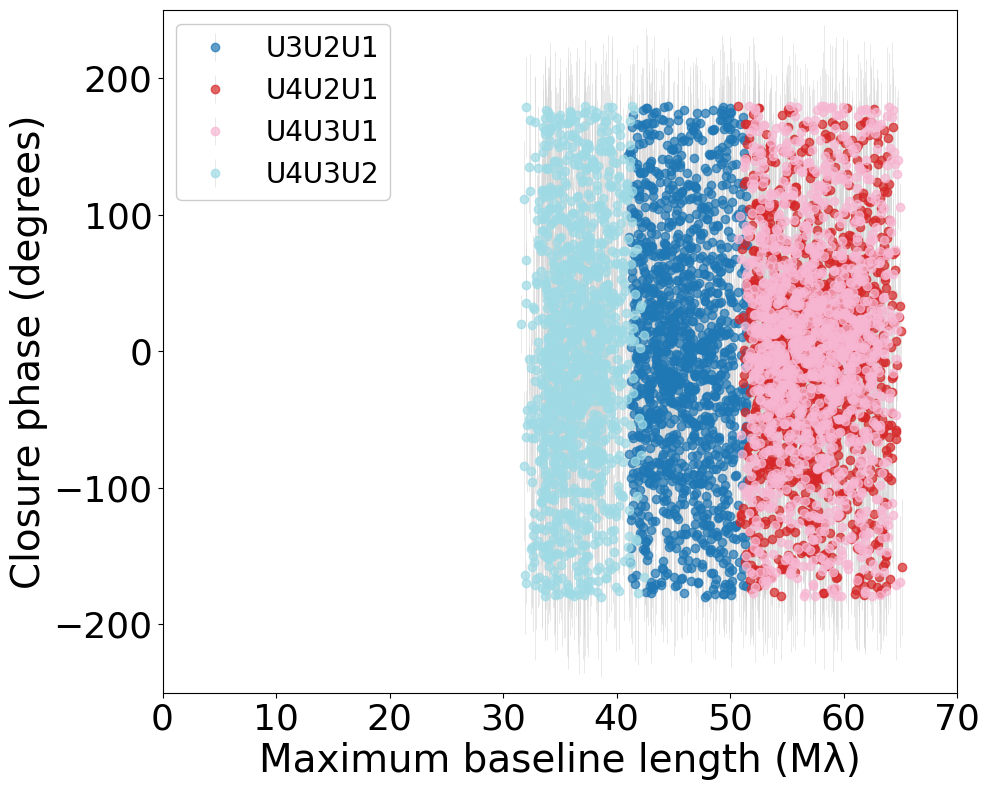}
      \caption{Top: Data of the fringe tracker 2MASS~16042987$-$4441336. Bottom: Data of the target PMN~J1604$-$4441.
      Left: UV coverage. Middle: Squared visibilities as a function of baseline length. Right: Closure phase as a function of maximum baseline of the telescope triangle. In all plots, the colors indicate different telescope pairs or triangles as shown in the legends.
              }
         \label{fig:data_J1604}
   \end{figure*}

PMN~J1604$-$4441 (Fig.~\ref{fig:data_J1604}) is an AGN of unknown type observed in the radio band within the TANAMI program \citep{muller18}. We observed it with GRAVITY in February 2024 along with a fringe tracker star 2MASS~16042987$-$4441336 separated by 12$\arcsec$ from it. According to 2MASS, the fringe tracker has a K-band magnitude at the limit for GRAVITY observations, which likely explains the noisier data as compared to 2MASS~J13285936$-$5608191. Nevertheless, possibly due to the smaller separation between the fringe tracker and the target, the data of the target itself are less noisy, especially when ignoring the shortest and longest wavelength bins, but they are still very low in visibility amplitudes. Therefore, no detailed modeling on the data could be performed.

\FloatBarrier
\subsection{TXS~1811+062}\label{app1811}
\begin{figure*}[h]
            \includegraphics[width=0.33\textwidth]{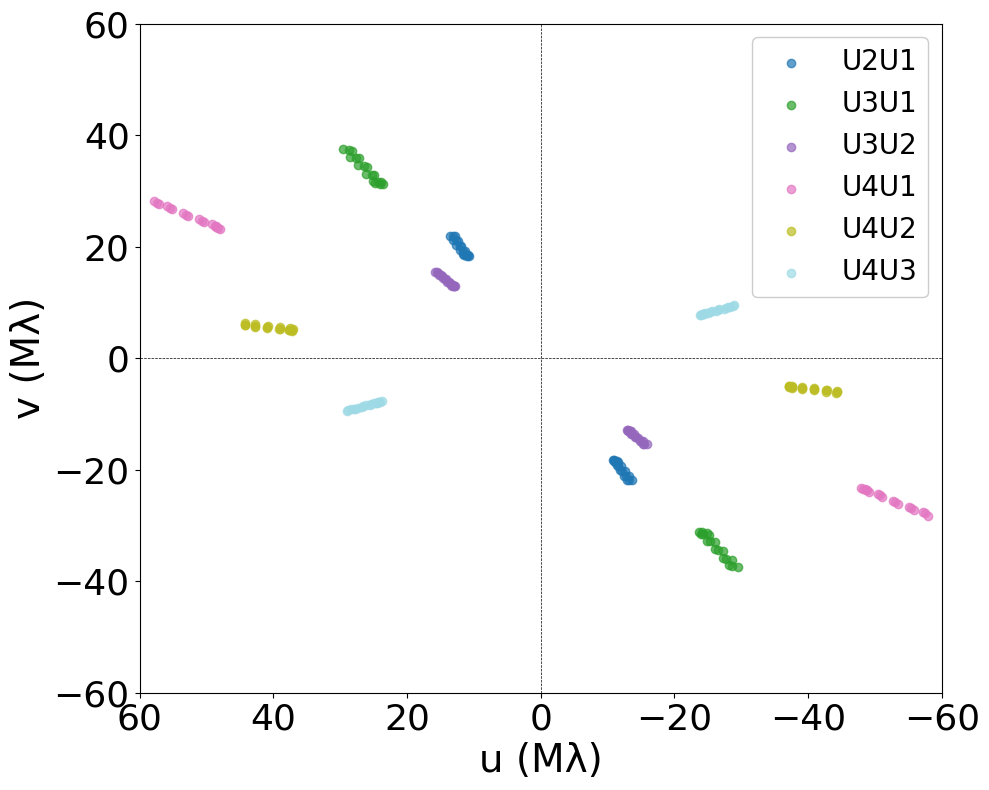}\includegraphics[width=0.33\textwidth]{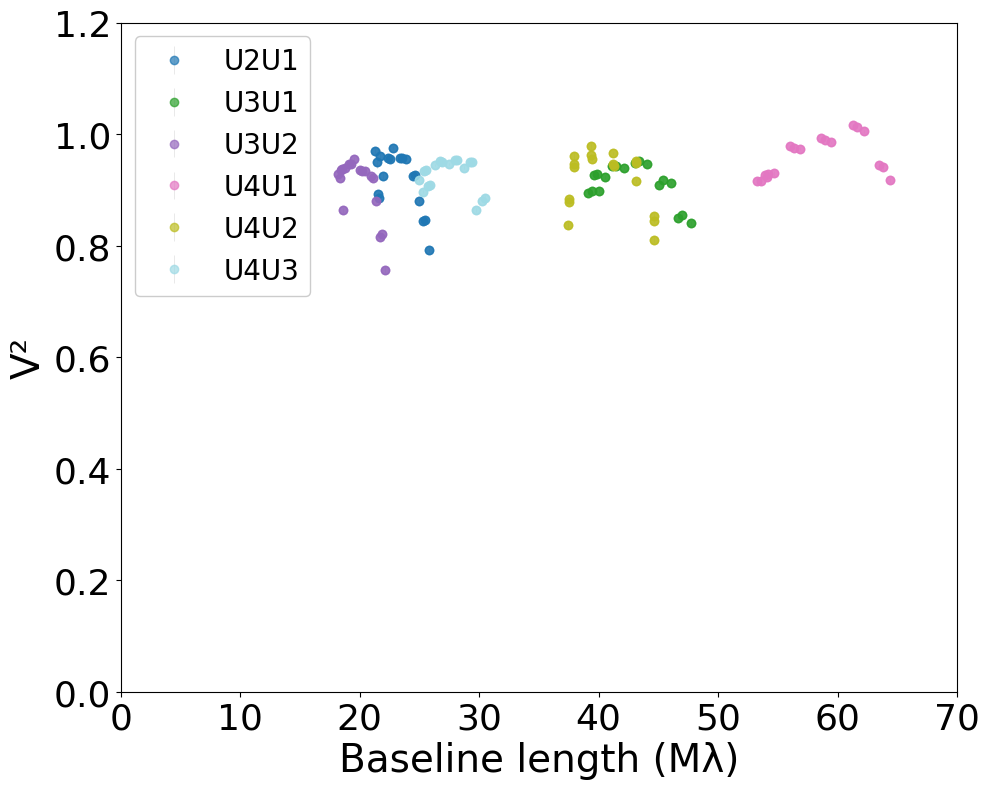}\includegraphics[width=0.33\textwidth]{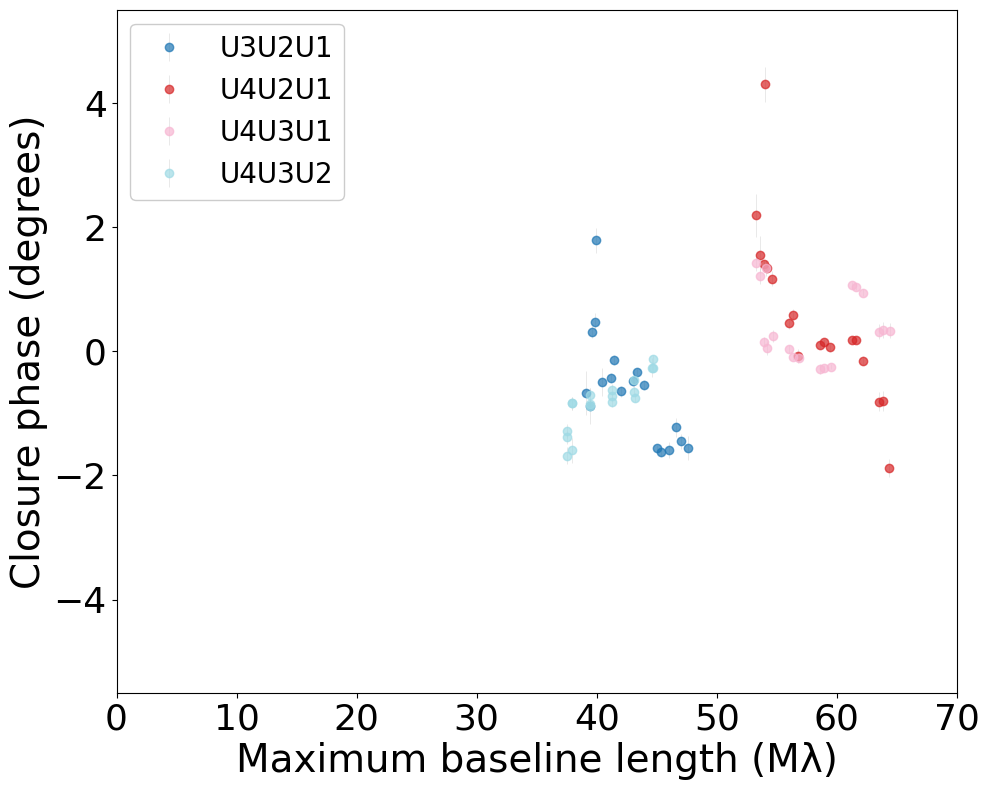}\\
            \includegraphics[width=0.33\textwidth]{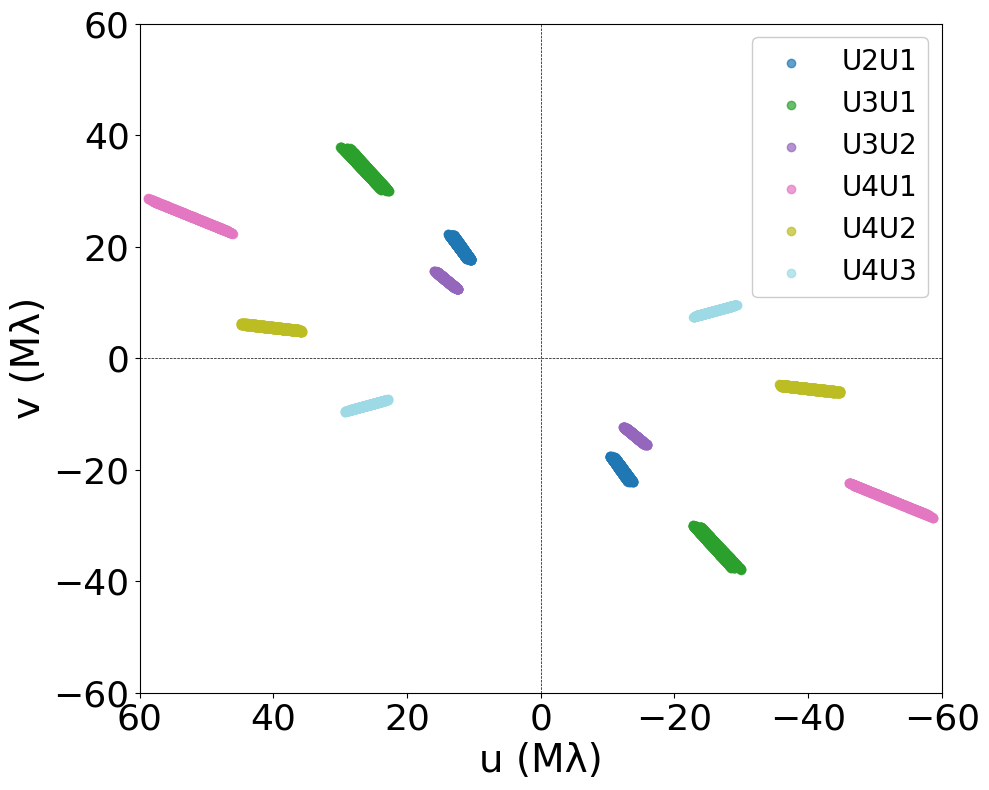}\includegraphics[width=0.33\textwidth]{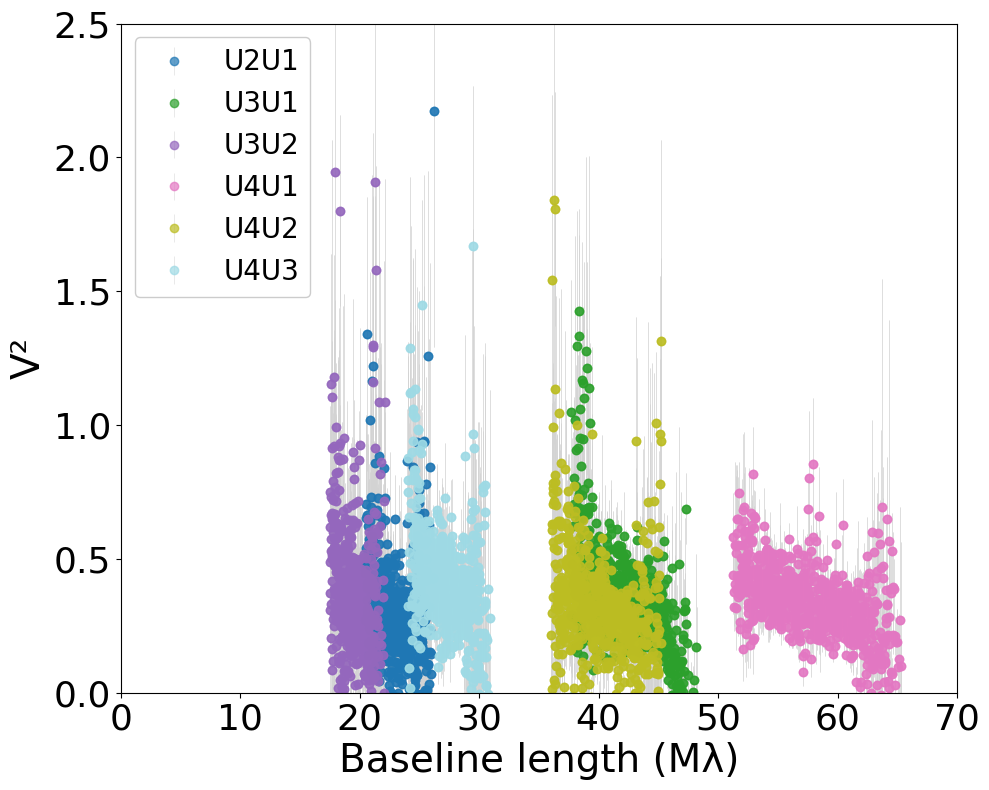}\includegraphics[width=0.33\textwidth]{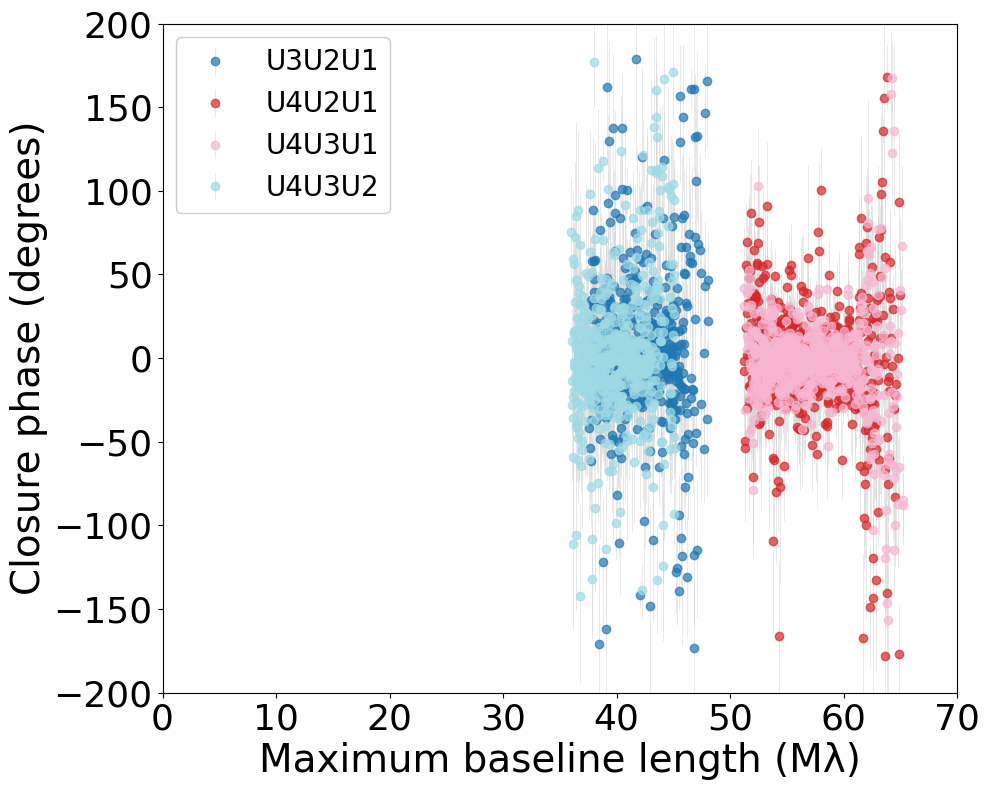}
      \caption{Top: Data of the fringe tracker 2MASS~J18133461+0615517 taken on July 3, 2023. Bottom: Data of the target TXS~1811+062.
      Left: UV coverage. Middle: Squared visibilities as a function of baseline length. Right: Closure phase as a function of maximum baseline of the telescope triangle. In all plots, the colors indicate different telescope pairs or triangles as shown in the legends.
              }
         \label{fig:data_1811_Jul23}
   \end{figure*}

TXS~1811+062 is a BL~Lac type object observed in radio within the MOJAVE program \citep{lister18}. It was observed three times with GRAVITY along with its fringe tracker star 2MASS~J18133461+0615517 separated by 20.4$\arcsec$. The first observations in July 2023 were interrupted after the first execution of the target because of technical problems with one of the telescopes. Nevertheless, we show the data from this observation in Fig.~\ref{fig:data_1811_Jul23}. Although the visibility amplitudes of the target are low, the target is clearly detected with much smaller uncertainties than in the other targets. However, due to the relatively large scatter, we did not attempt any detailed modeling of these data either.

\begin{figure*}[h]
            \includegraphics[width=0.33\textwidth]{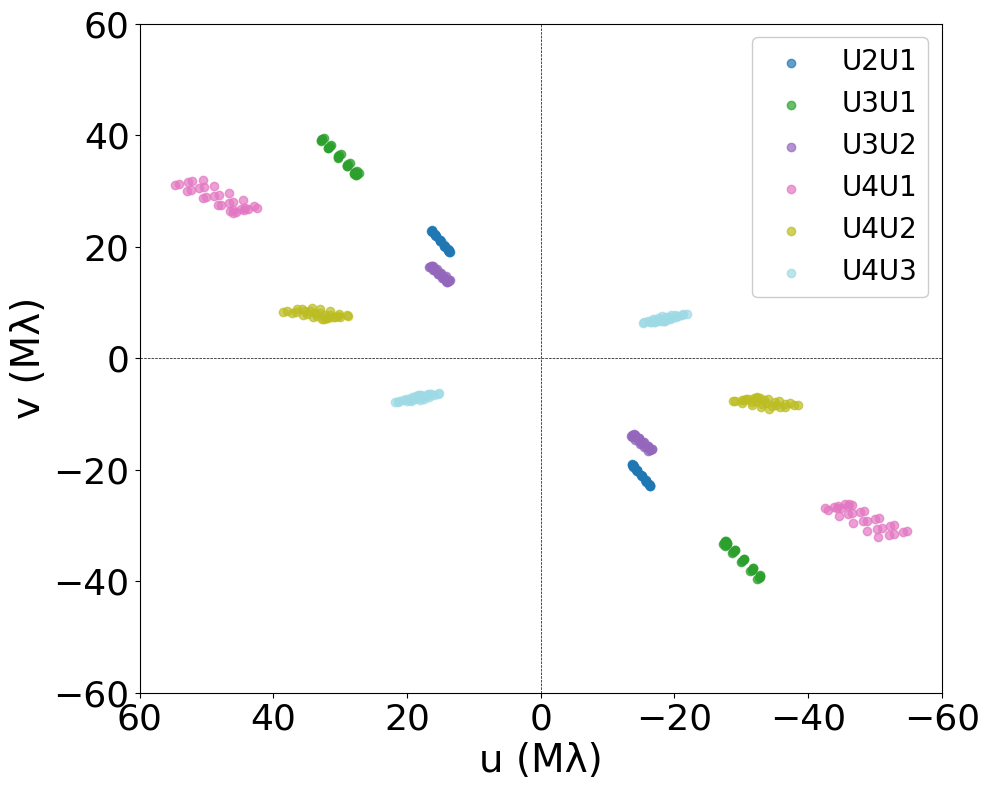}\includegraphics[width=0.33\textwidth]{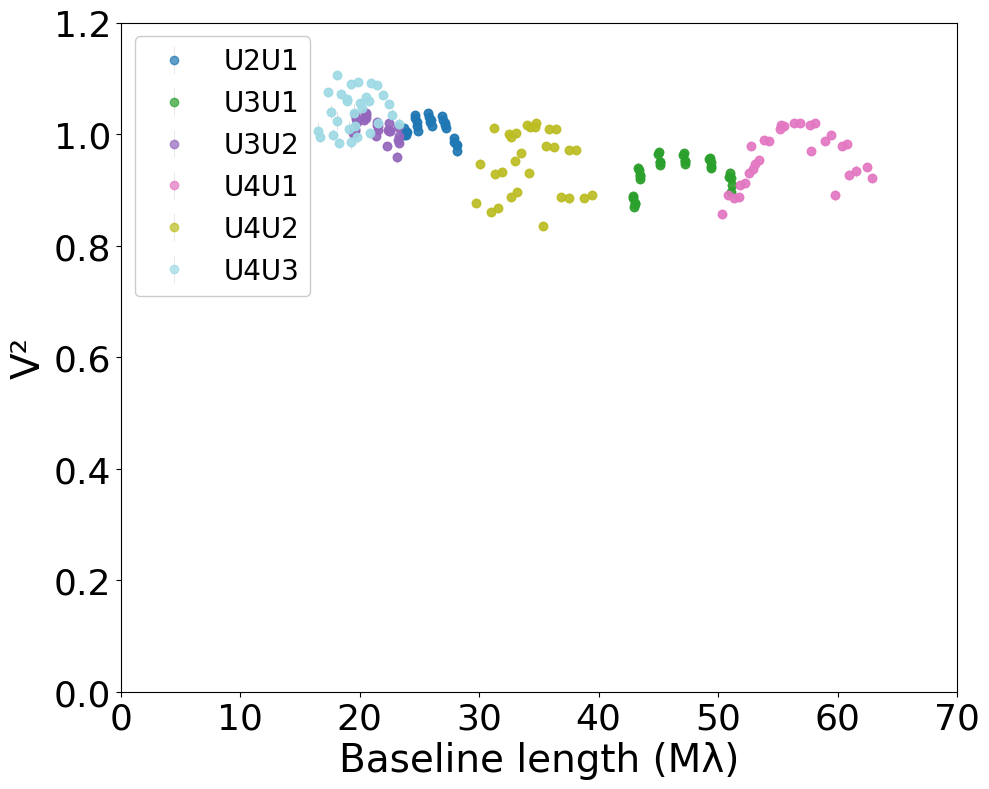}\includegraphics[width=0.33\textwidth]{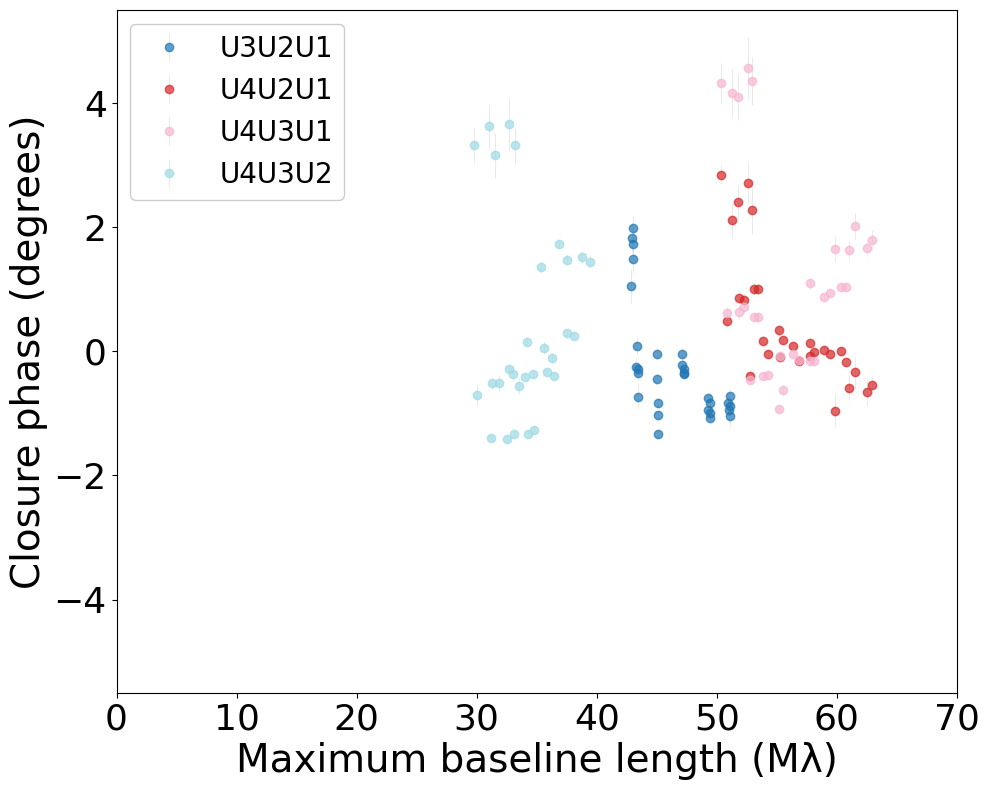}\\
            \includegraphics[width=0.33\textwidth]{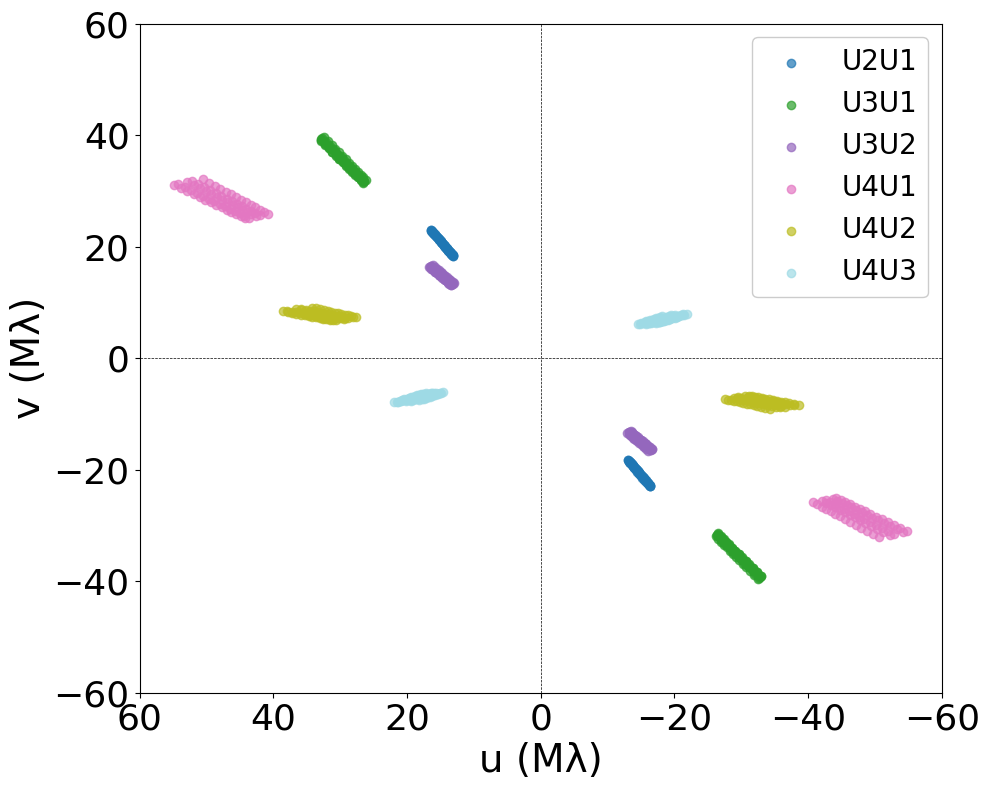}\includegraphics[width=0.33\textwidth]{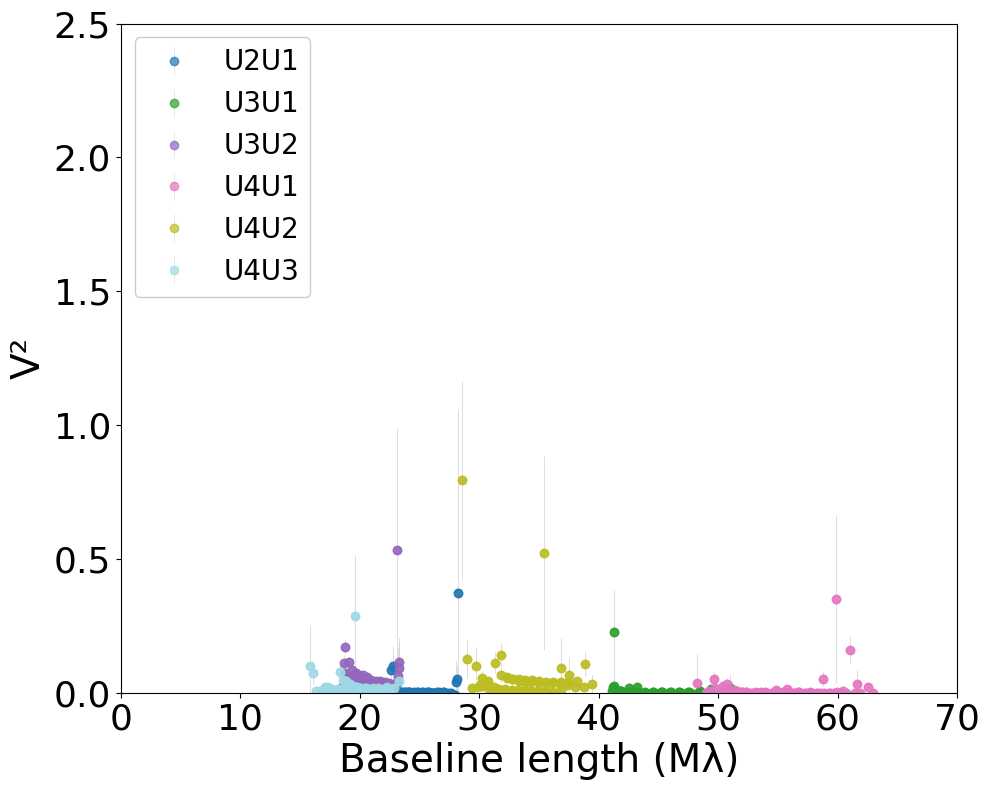}\includegraphics[width=0.33\textwidth]{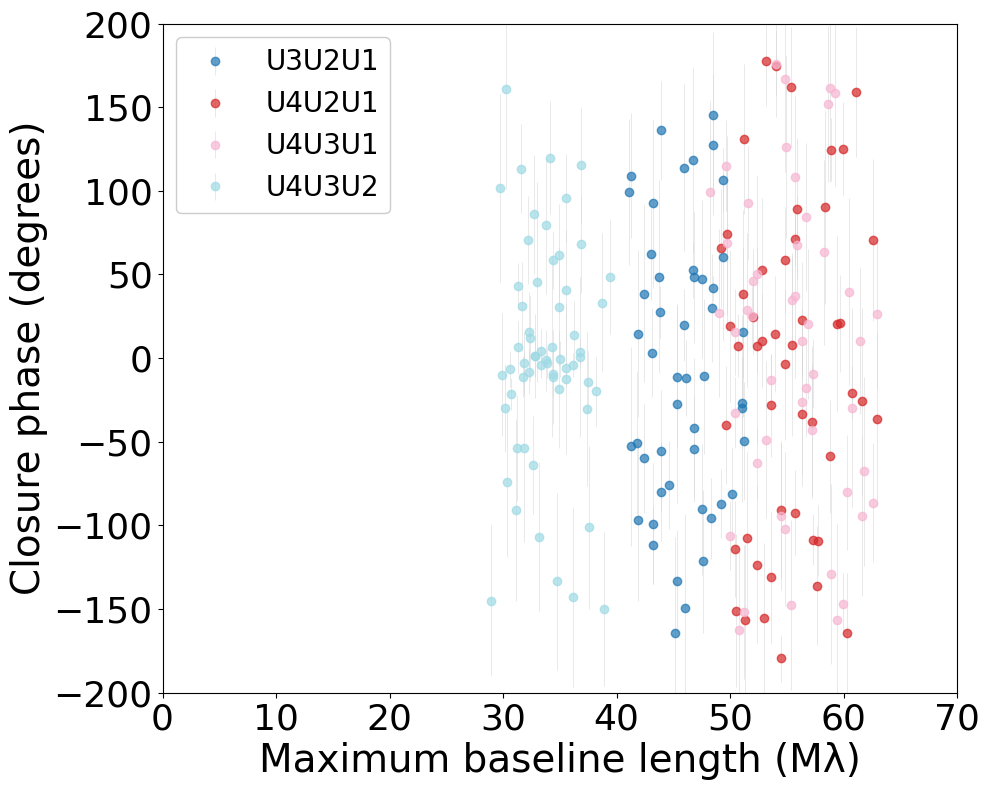}
      \caption{Top: Data of the fringe tracker 2MASS~J18133461+0615517 taken on July 11, 2025. Bottom: Data of the target TXS~1811+062.
      Left: UV coverage. Middle: Squared visibilities as a function of baseline length. Right: Closure phase as a function of maximum baseline of the telescope triangle. In all plots, the colors indicate different telescope pairs or triangles as shown in the legends.
              }
         \label{fig:data_1811_Jul25}
   \end{figure*}

\begin{figure*}[h]
            \includegraphics[width=0.33\textwidth]{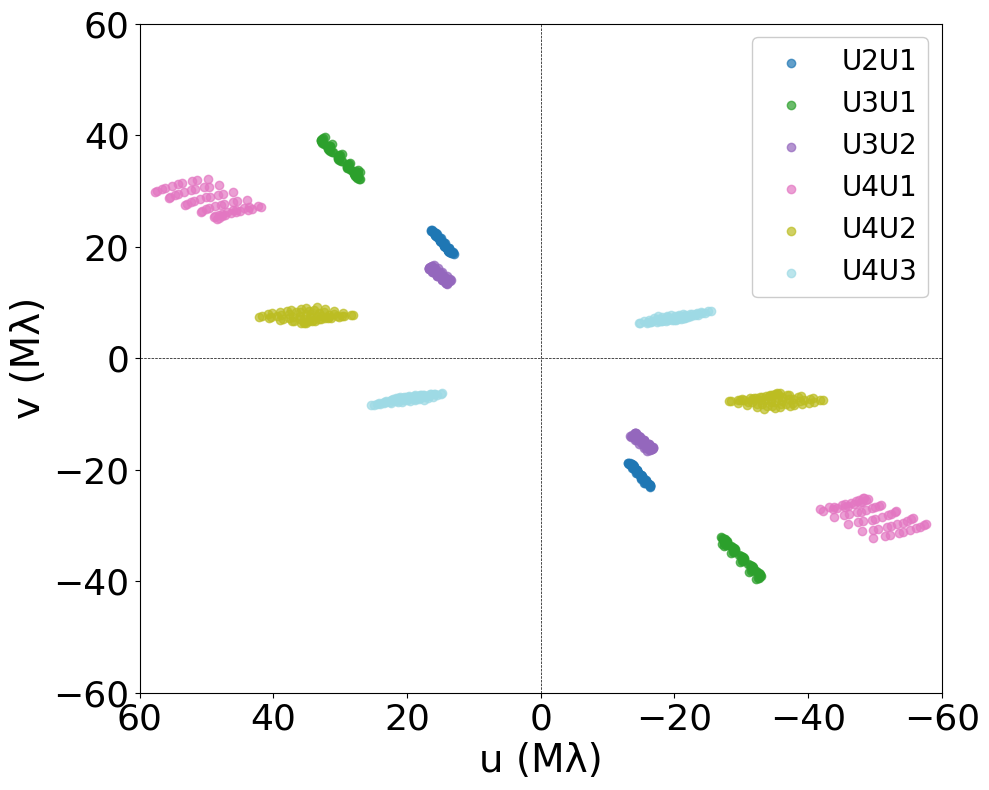}\includegraphics[width=0.33\textwidth]{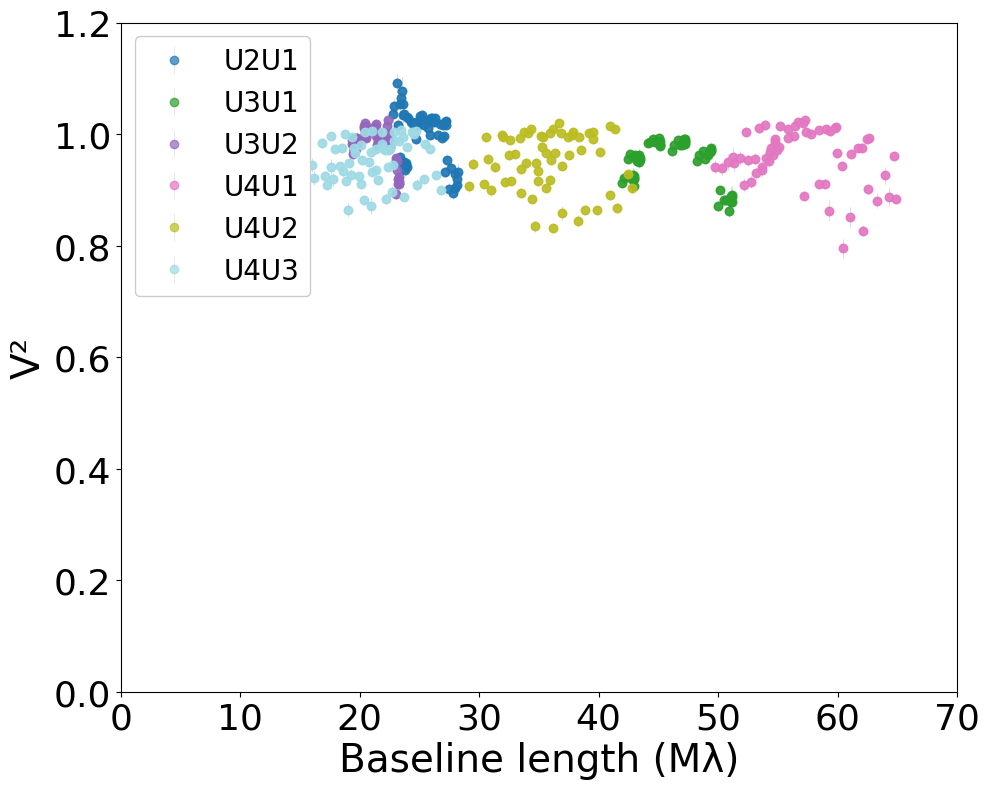}\includegraphics[width=0.33\textwidth]{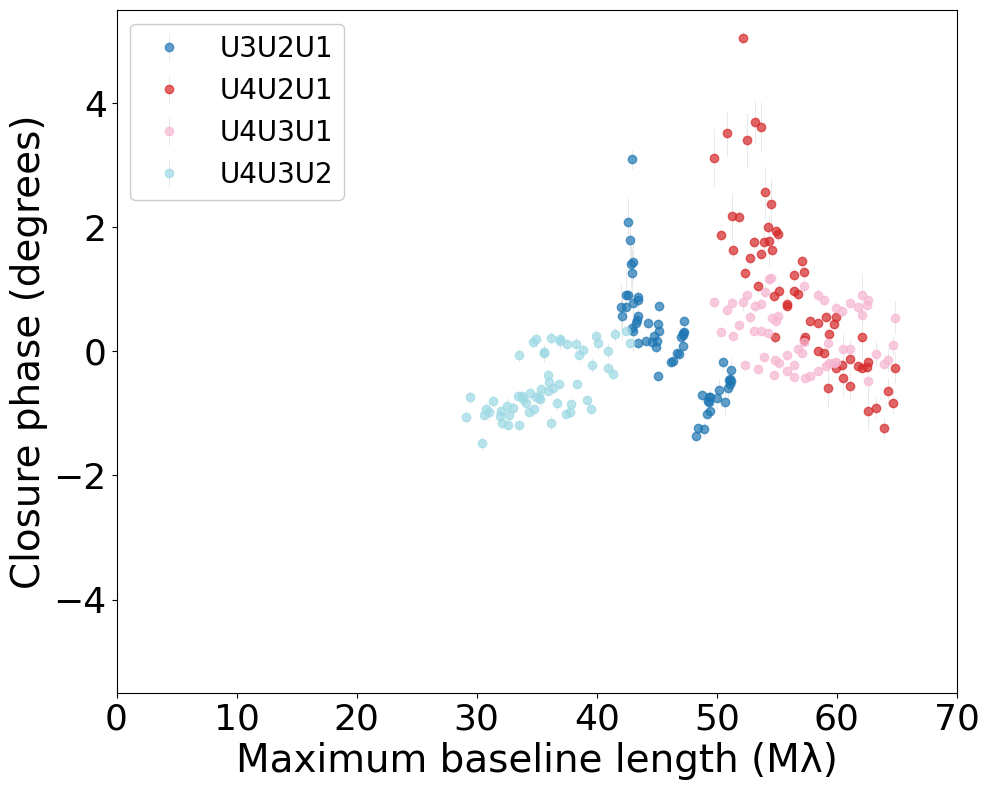}\\
            \includegraphics[width=0.33\textwidth]{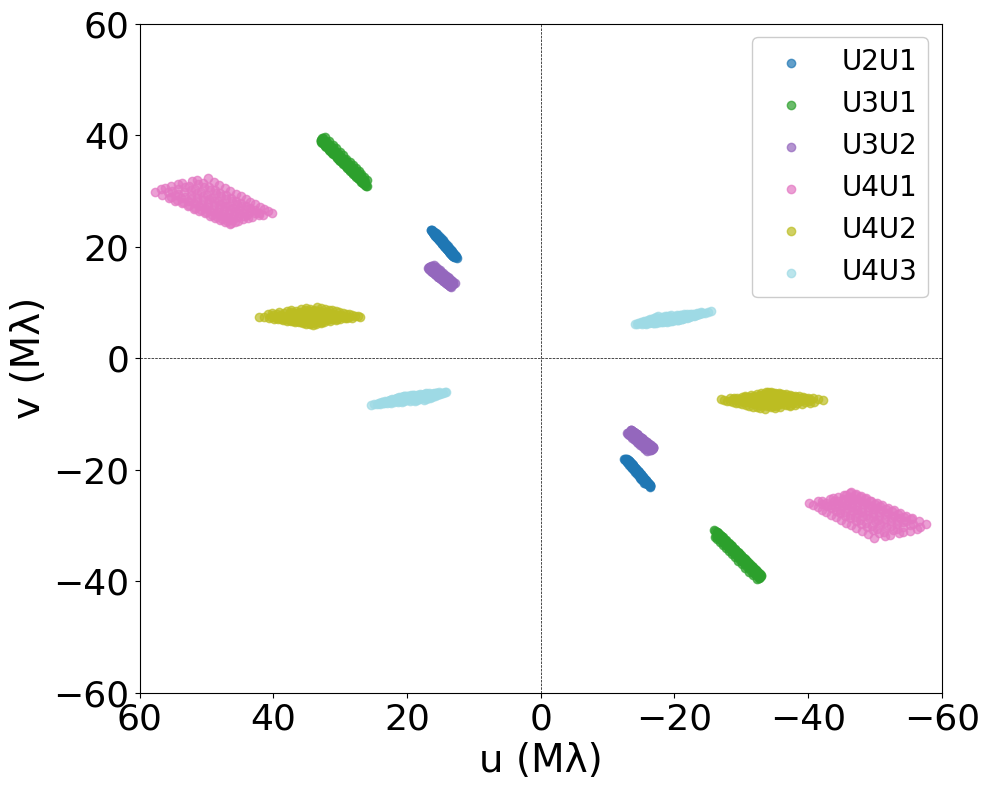}\includegraphics[width=0.33\textwidth]{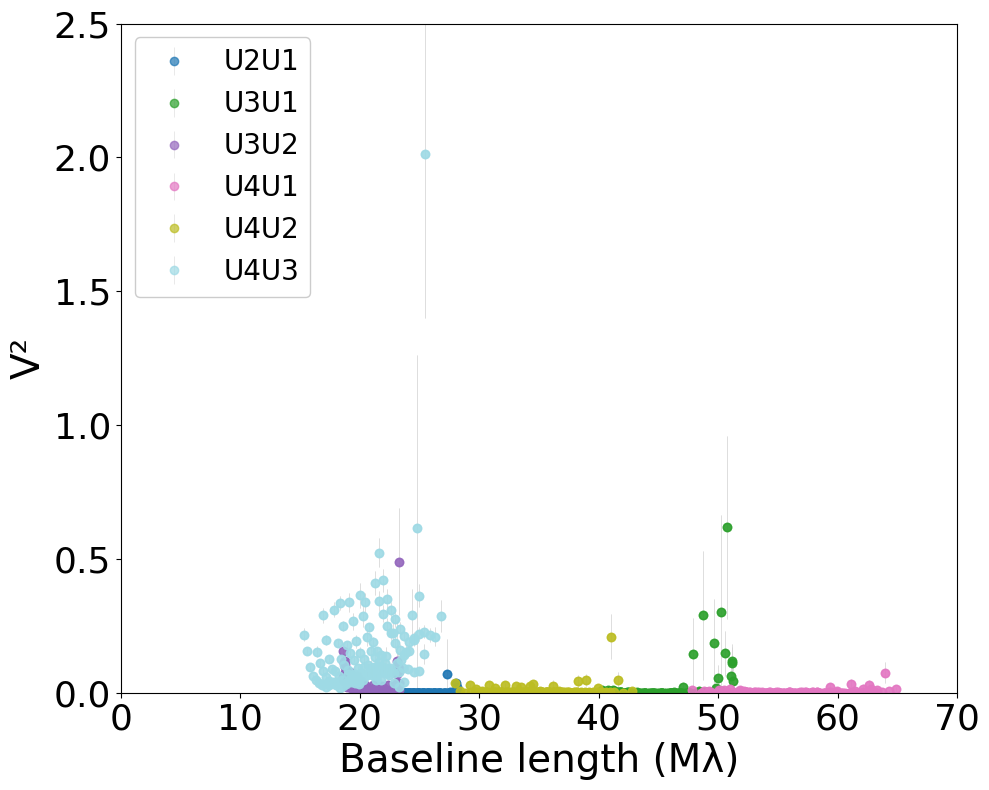}\includegraphics[width=0.33\textwidth]{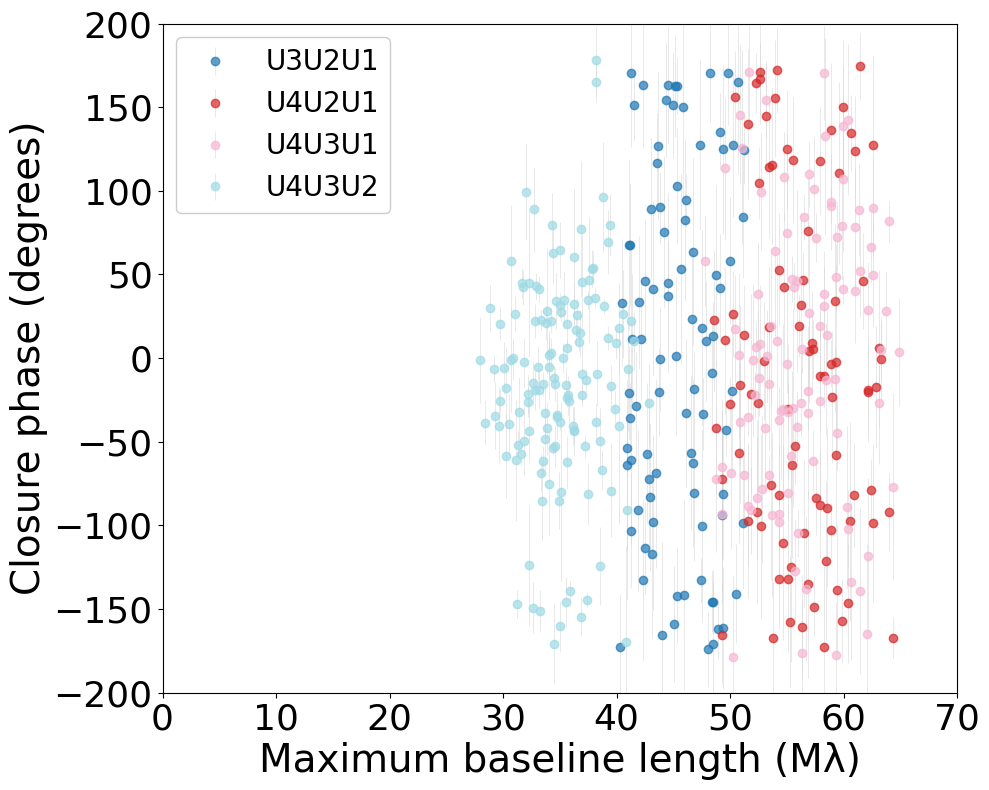}
      \caption{Top: Data of the fringe tracker 2MASS~J18133461+0615517 taken on August 6, 2025. Bottom: Data of the target TXS~1811+062.
      Left: UV coverage. Middle: Squared visibilities as a function of baseline length. Right: Closure phase as a function of maximum baseline of the telescope triangle. In all plots, the colors indicate different telescope pairs or triangles as shown in the legends.
              }
         \label{fig:data_1811_Aug25}
   \end{figure*}

Because there was a clear detection in these data, we proposed for a new program to re-observe the source in the low spectral resolution, which in principle should give a better signal-to-noise ratio if the conditions are otherwise similar. These observations were taken in July and August 2025 and we show the data in Figs.~\ref{fig:data_1811_Jul25} and \ref{fig:data_1811_Aug25}. Quite surprisingly, the source is not detected with the visibility amplitudes being very low and closure phases corresponding to pure noise. We checked if there was a difference in the observing conditions between the 2023 and 2025 observations. The seeing in all executions has been similar and within the requested constraints of $< 0.6\arcsec$. The coherence time in July 2023 was between $7-9$\,ms, while in July 2025 it was somewhat lower between $4-7$\,ms but again higher in August 2025 where it was between $6-15$\,ms. As explained in \cite{gravity_wide_22}, from the atmospheric conditions, the isoplanatic angle affects the wide-field observations the most with a larger value resulting in smaller losses. In both July 2023 and August 2025 the angle was around $3\arcsec$ while in July 2025 it was somewhat lower between $1.6-2.4\arcsec$, but still similar enough that it cannot explain the non-detection when looking at the predicted coherence loss from \cite{gravity_wide_22}. Therefore, the overall observing conditions have been similar enough so that they cannot explain the non-detection in 2025. 

As the targets are known to be variable, we also checked if the brightness of the target had changed between 2023 and 2025. For this we utilized the Asteroid Terrestrial-impact Last Alert System (ATLAS\footnote{\href{https://fallingstar-data.com/}{https://fallingstar-data.com/}}) all-sky survey data \citep{tonry2018}, which were available for both times. The data are taken with the orange filter (578\,nm) and we do not know the exact color transformation between it and the near-infrared K-band but this will give us an indication if the source brightness had changed between the epochs. As the ATLAS data are quite noisy (see \citealt{kouch25} for the usage of ATLAS data in constructing light curves of blazars), we averaged observations taken within $\pm3.5$ days of our GRAVITY observations and discarded data points with uncertainties larger than 1 mag. Interestingly, the source has been {\it fainter} in 2023 with an average o-band magnitude of $19.1\pm0.2$ compared to July 2025 ($18.6\pm0.04$) and August 2025 ($18.4\pm0.05$). 

Finally, the coherence length $\lambda^2/\Delta\lambda$ is significantly shorter in low spectral resolution ($R\sim 20$) used in 2025 compared to medium spectral resolution ($R\sim 500$) used in 2023. This leads to an increased likelihood to miss the fringe packet in low spectral resolution owing to uncertainties in the coordinates of the fringe tracking and science targets, as well as internal calibration uncertainties within the interferometer. Indeed, we found a difference of ESO P2 tool and Gaia coordinates of the fringe tracking star of 64 mas, which would be significant compared to the coherence length in low resolution, but not in medium resolution. This may explain the lack of detection in our 2025 data in low spectral resolution.

\subsection{QSO~B2201+1711}
\begin{figure*}[h]
            \includegraphics[width=0.33\textwidth]{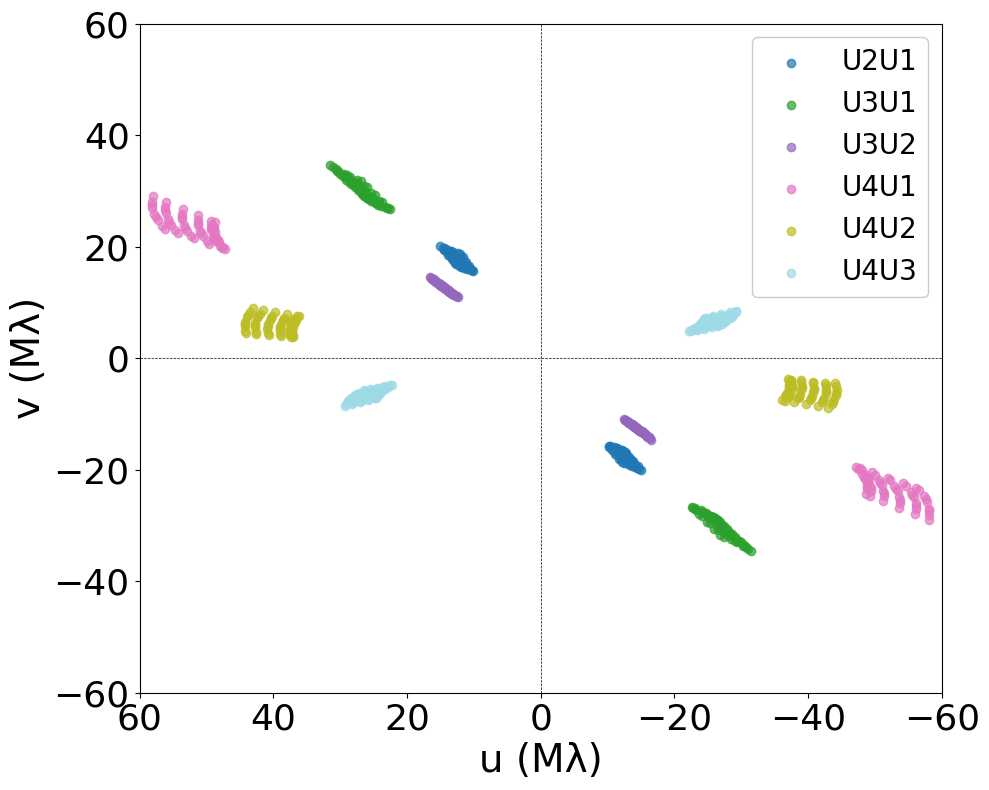}\includegraphics[width=0.33\textwidth]{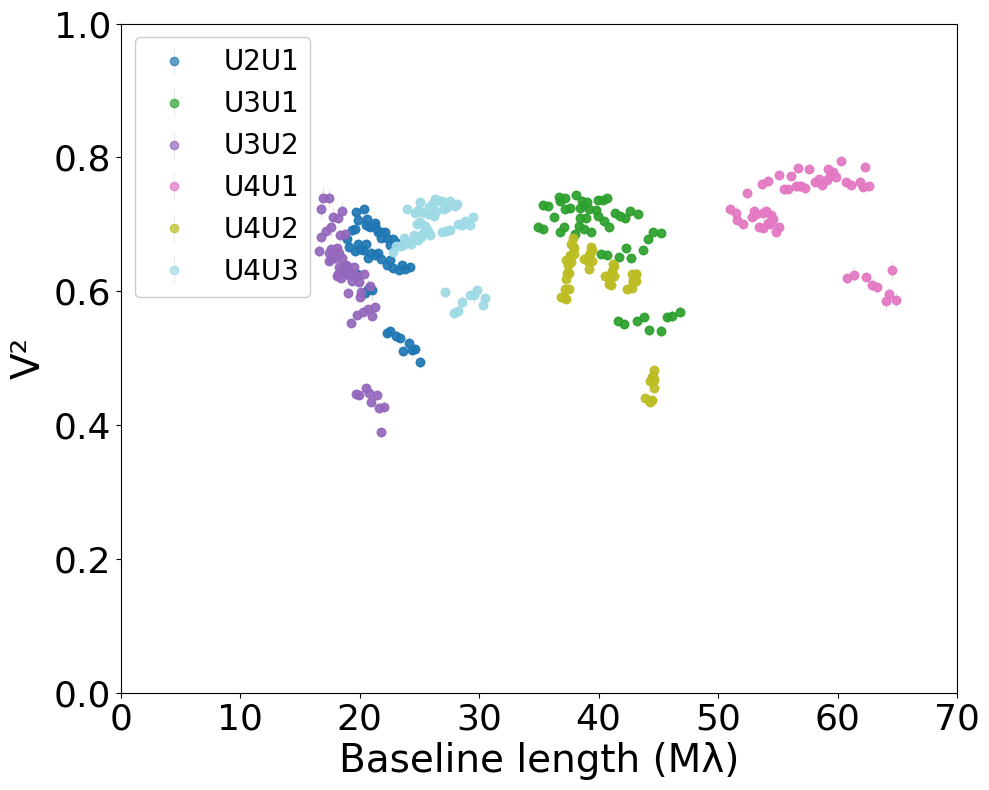}\includegraphics[width=0.33\textwidth]{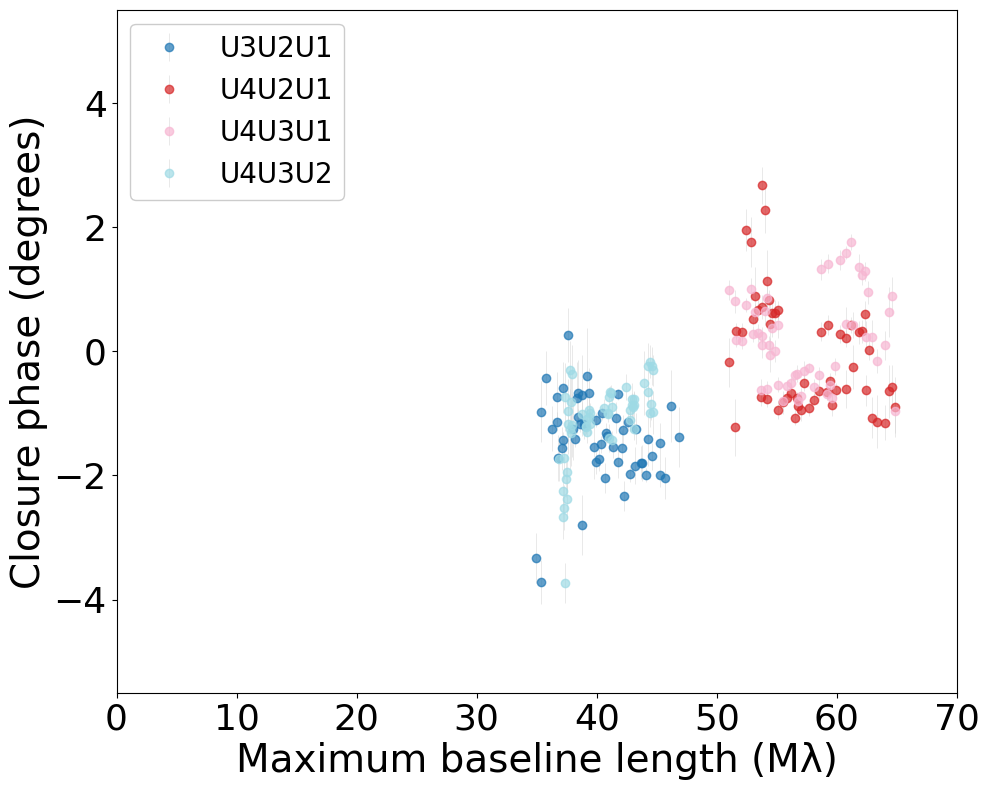}\\
            \includegraphics[width=0.33\textwidth]{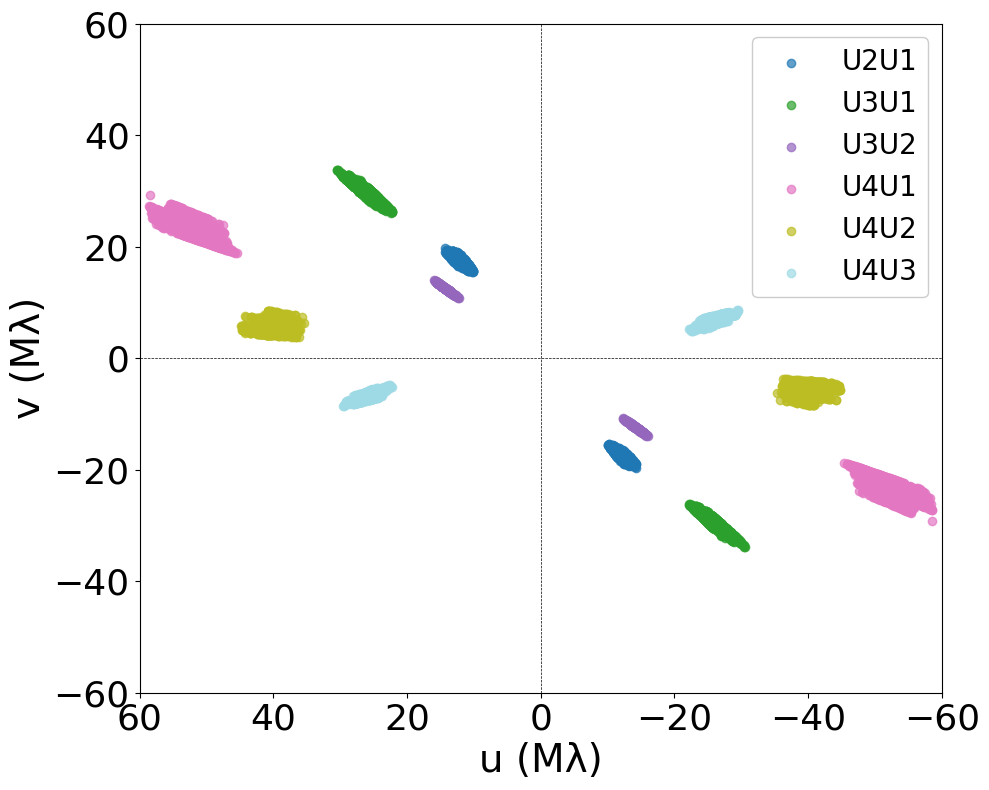}\includegraphics[width=0.33\textwidth]{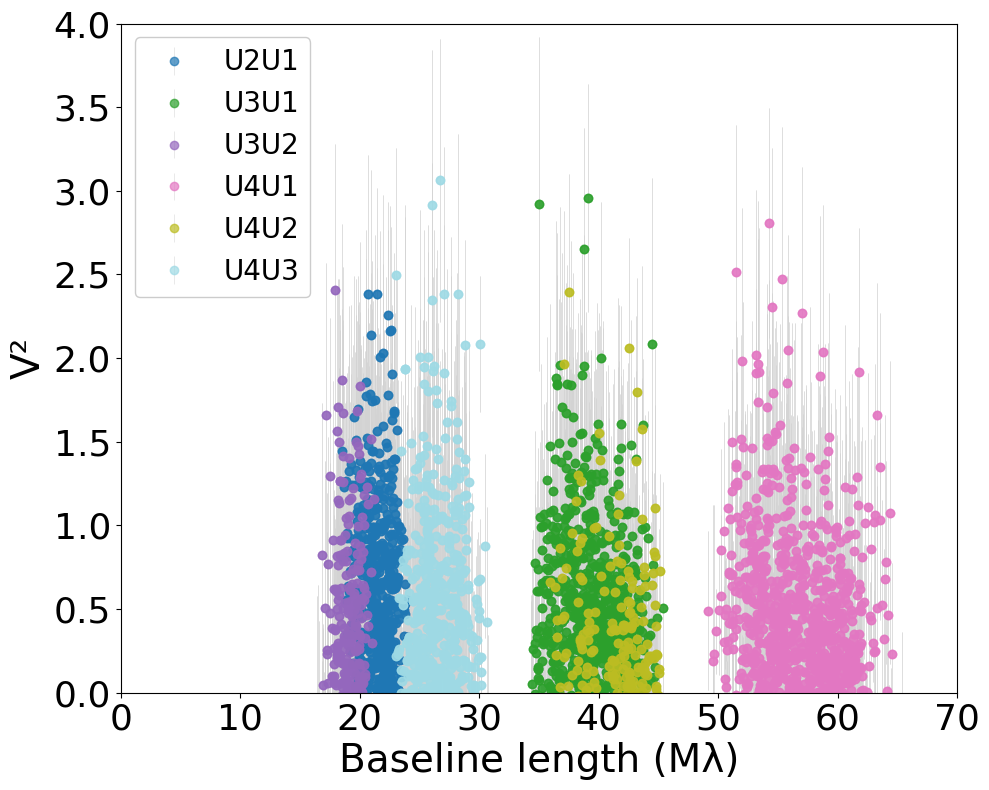}\includegraphics[width=0.33\textwidth]{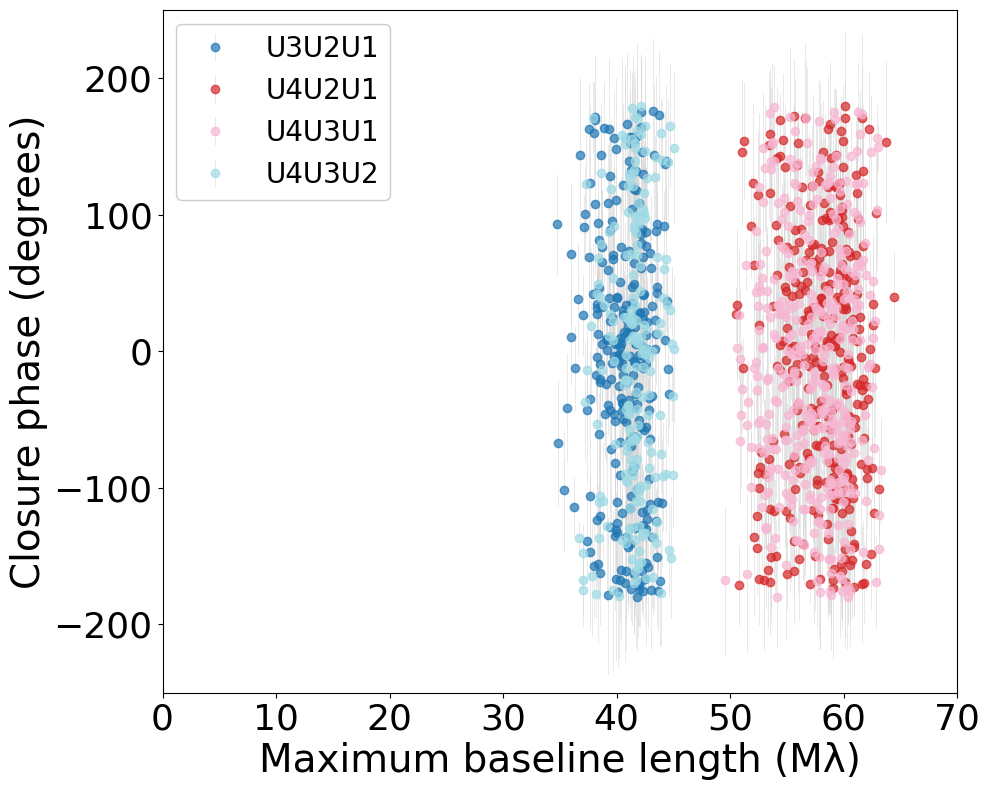}
      \caption{Top: Data of the fringe tracker GPM~330.860278+17.426672 taken on July 1, 2023. Bottom: Data of the target QSO~B2201+1711.
      Left: UV coverage. Middle: Squared visibilities as a function of baseline length. Right: Closure phase as a function of maximum baseline of the telescope triangle. In all plots, the colors indicate different telescope pairs or triangles as shown in the legends.
              }
         \label{fig:data_2201_Jul1}
   \end{figure*}

\begin{figure*}[h]
            \includegraphics[width=0.33\textwidth]{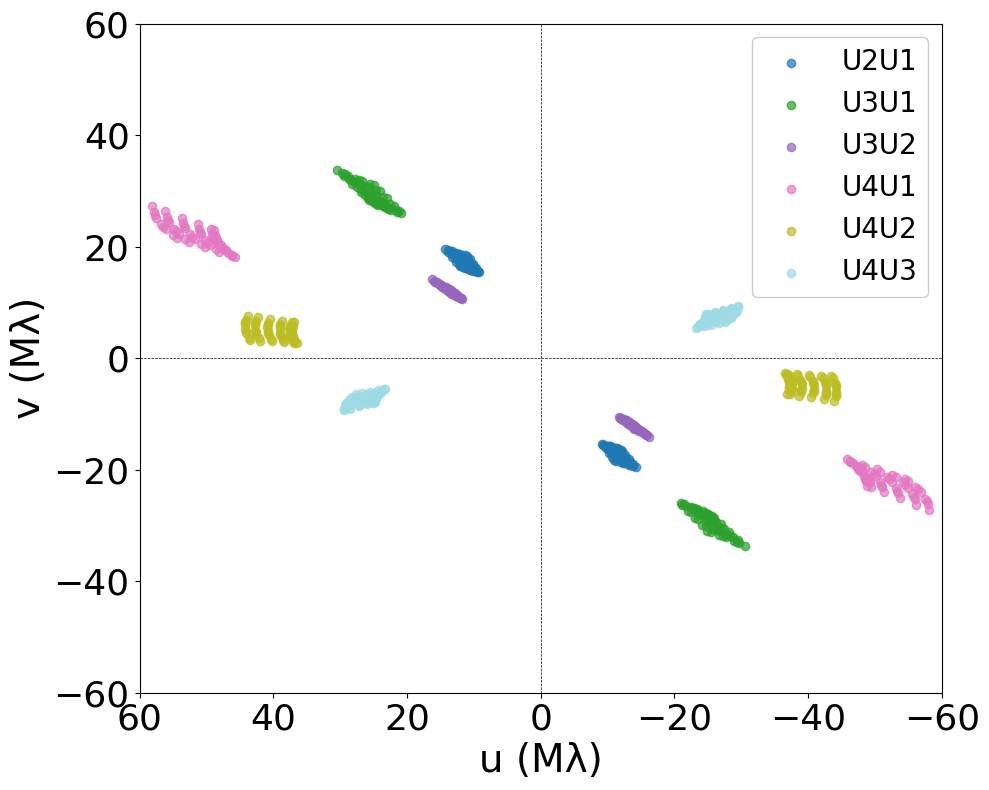}\includegraphics[width=0.33\textwidth]{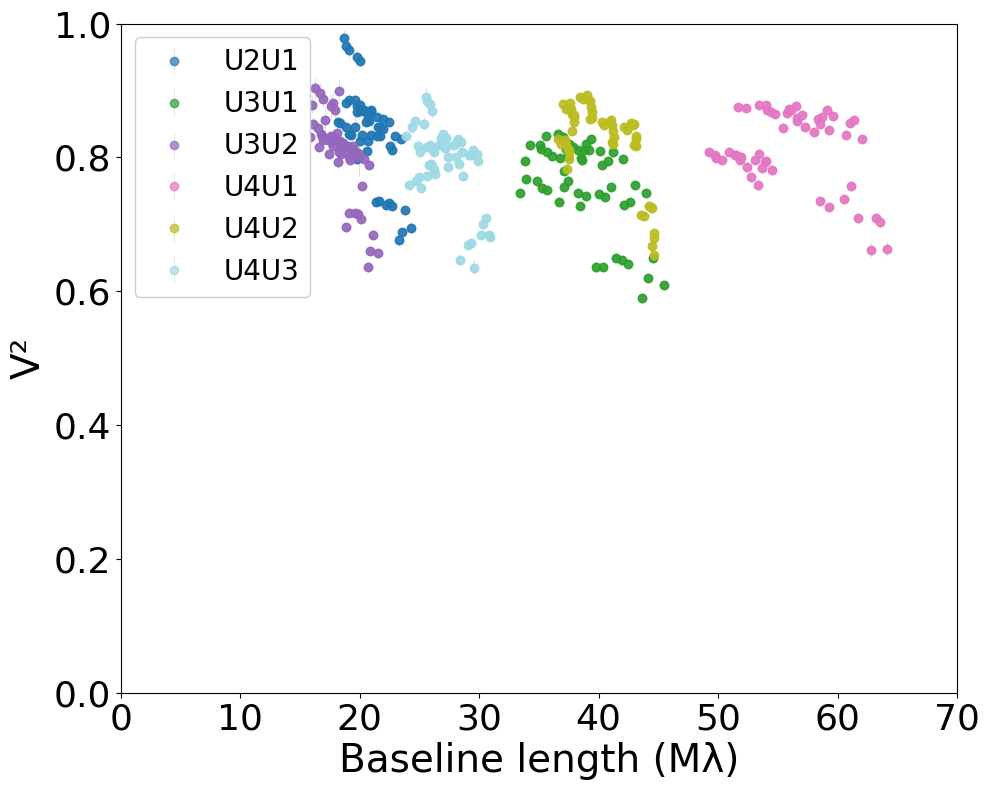}\includegraphics[width=0.33\textwidth]{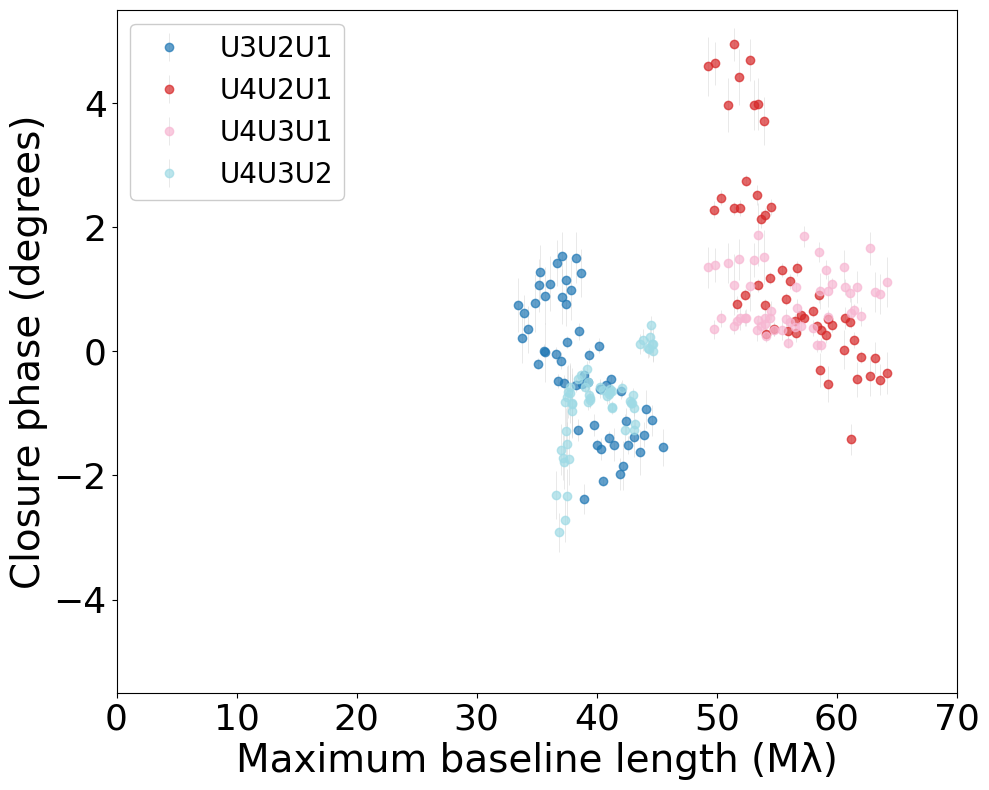}\\
            \includegraphics[width=0.33\textwidth]{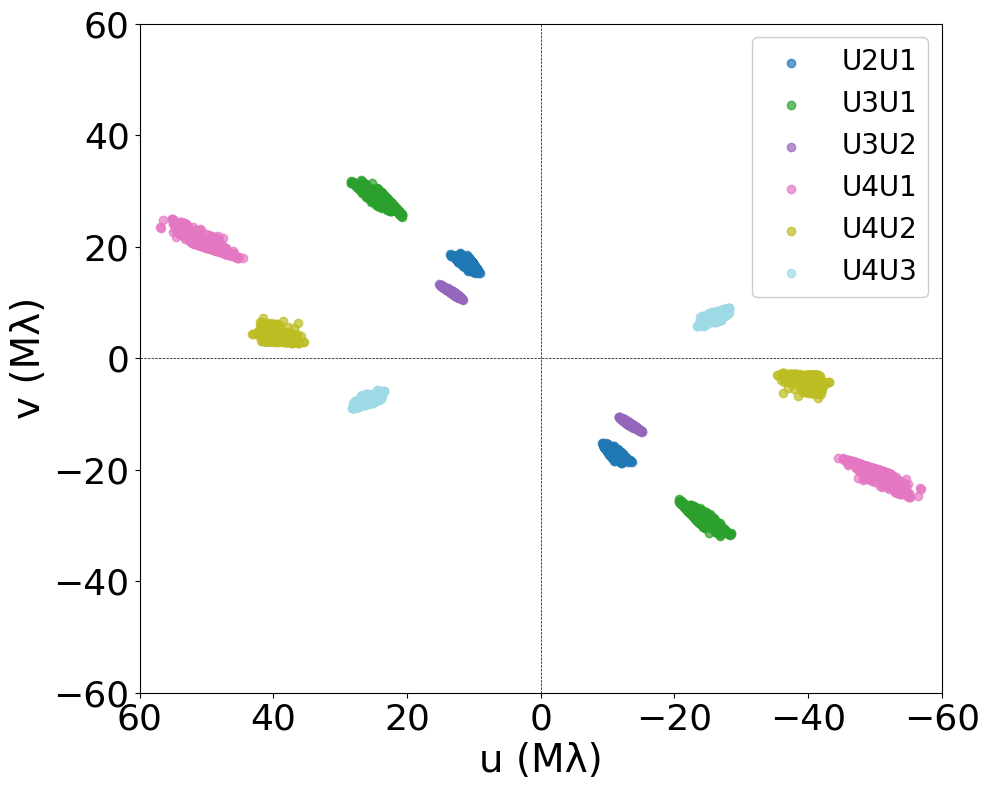}\includegraphics[width=0.33\textwidth]{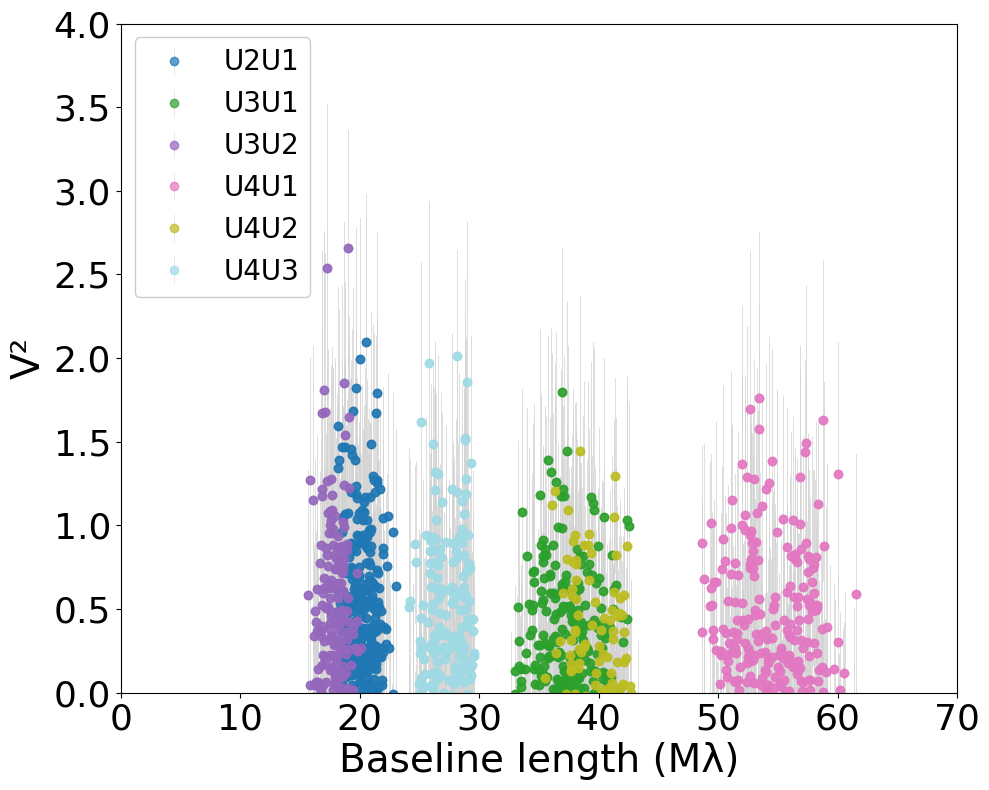}\includegraphics[width=0.33\textwidth]{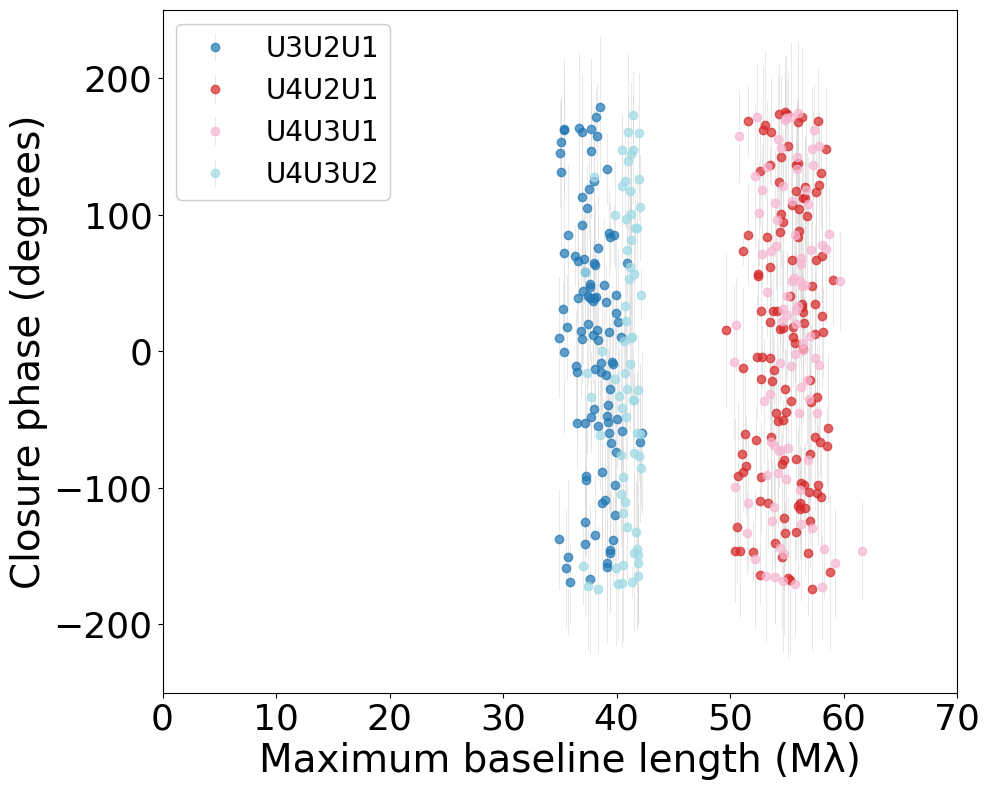}
      \caption{Top: Data of the fringe tracker GPM~330.860278+17.426672 taken on July 6, 2023. Bottom: Data of the target QSO~B2201+1711.
      Left: UV coverage. Middle: Squared visibilities as a function of baseline length. Right: Closure phase as a function of maximum baseline of the telescope triangle. In all plots, the colors indicate different telescope pairs or triangles as shown in the legends.
              }
         \label{fig:data_2201_Jul6}
   \end{figure*}

QSO~B2201+1711 is a flat-spectrum radio quasar observed in radio within the MOJAVE program \citep{lister18}. It was observed in July 2023 on two days within the same GRAVITY sessions. The fringe-tracker star GPM~330.860278+17.426672 was separated by $14\arcsec$ from it. We show the data of the two executions separately in Figs.~\ref{fig:data_2201_Jul1} and \ref{fig:data_2201_Jul6}. Both executions were taken in very good conditions with a seeing of $0.5-0.7\arcsec$, while in the first execution the coherence time was better ($10-17$\,ms) than in the second one ($5-7$\,ms). This may explain the slightly higher values of the squared visibilities of the target in the first execution, but in both cases they are very noisy and the closure phases are consistent with noise so that we consider this a non-detection. 

\end{appendix}

\end{document}